\documentclass[useAMS,usenatbib]{mn2e}

\renewcommand\sfdefault{ptm}
\usepackage{sfmath}

\usepackage{lscape}
\usepackage{aas_macros}

\usepackage{epsfig}
\usepackage{natbib}

\newbox\grsign \setbox\grsign=\hbox{$>$} \newdimen\grdimen \grdimen=\ht\grsign
\newbox\simlessbox \newbox\simgreatbox \newbox\simpropbox \newbox\wtildebox 
\setbox\simgreatbox=\hbox{\raise.5ex\hbox{$>$}\llap
    {\lower.5ex\hbox{$\sim$}}}\ht1=\grdimen\dp1=0pt
\setbox\simlessbox=\hbox{\raise.5ex\hbox{$<$}\llap
    {\lower.5ex\hbox{$\sim$}}}\ht2=\grdimen\dp2=0pt

\newcommand{\be}{\mbox{\begin{equation}}}
\newcommand{\ee}{\mbox{\end{equation}}}

\newcommand{\Cref}{\mbox{$m_{\rm ref}$}}

\newcommand{\msun}{\mbox{M$_\odot$}}

\title{Formation versus destruction: the evolution of the star cluster population in galaxy mergers}   
\author{J.~M.~Diederik Kruijssen,$^{1,2,3}$\thanks{kruijssen@mpa-garching.mpg.de} F.~Inti~Pelupessy,$^3$ Henny~J.~G.~L.~M.~Lamers,$^2$ \newauthor Simon~F.~Portegies~Zwart,$^3$ Nate Bastian$^{4}$ and Vincent~Icke$^3$\\
$^1$Max-Planck Institut f\"{u}r Astrophysik, Karl-Schwarzschild-Stra\ss e 1, 85748, Garching, Germany\\
$^2$Astronomical Institute, Utrecht University, PO Box 80000, 3508 TA Utrecht, The Netherlands\\
$^3$Leiden Observatory, Leiden University, PO Box 9513, 2300 RA Leiden, The Netherlands\\
$^4$Excellence Cluster Universe, Technische Universit\"{a}t M\"{u}nchen, Boltzmannstra\ss e 2, 85748 Garching, Germany}

\begin{document}

\date{Accepted 2011 December 1. Received 2011 December 1; in original form 2011 November 11.}

\pagerange{\pageref{firstpage}--\pageref{lastpage}} \pubyear{2011}
\label{firstpage}

\maketitle

\begin{abstract}
{Interacting galaxies are well-known for their high star formation rates and rich star cluster populations, but it is also recognized that the rapidly changing tidal field can efficiently destroy clusters. We use numerical simulations of merging disc galaxies to investigate which mechanism dominates. The simulations include a model for the formation and evolution of the entire star cluster population, accounting for the evaporation of clusters due to two-body relaxation and tidal shocks. We find that the dynamical heating of stellar clusters by tidal shocks is about an order of magnitude higher in interacting galaxies than in isolated galaxies. This is driven by the increased gas density, and is sufficient to destroy star clusters at a higher rate than new clusters are formed: the total number of stellar clusters in the merger remnant is 2--50\% of the amount in the progenitor discs, with low-mass clusters being disrupted preferentially. By adopting observationally motivated selection criteria, we find that the observed surplus of star clusters in nearby merging galaxies with respect to isolated systems is caused by the observational bias to detect young, massive clusters, and marks a transient phase in galaxy evolution. We provide a general expression for the survival fraction of clusters, which increases with the gas depletion time-scale, reflecting that both the formation and the destruction of clusters are driven by the growth of the gas density. Due to the preferential disruption of low-mass clusters, the mass distribution of the surviving star clusters in a merger remnant develops a peak at a mass of about $10^{3}~\msun$, which evolves to higher masses at a rate of 0.3--0.4~dex per Gyr. Briefly after a merger, the peak mass depends weakly on the galactocentric radius, but this correlation disappears as the system ages due to the destruction of clusters on eccentric orbits. We discuss the similarities between the cluster populations of the simulated merger remnants and (young) globular cluster systems. Our results suggest that the combination of cluster formation and destruction should be widespread in the dense star-forming environments at high redshifts, which could provide a natural origin to present-day globular cluster systems.}
\end{abstract}


\setcounter{footnote}{0}

\section{Introduction} \label{sec:intro}
Merging and interacting galaxies host huge starbursts and large populations of young massive stellar clusters \citep[e.g.][]{holtzman92,schweizer96,whitmore99}. A galaxy interaction triggers inflows of interstellar gas towards the galaxy centres, where it fuels a burst of star formation \citep{hernquist89,mihos96,barnes96}. Merger-induced starbursts play a central role in the history of the universe, as galaxies are thought to have formed through hierarchical merging \citep[e.g.][]{white78,white91,cole00}. Some fraction of this star formation takes place in compact stellar clusters \citep{elmegreen83,whitmore99,bressert10} with masses in the range $10^2$--$10^8~\msun$ \citep{portegieszwart10}. The clusters that remain after a merger are often used as fossils to trace the formation history of the galaxy \citep{larsen01}. 

During the past two decades, observations with the Hubble Space Telescope have revealed that many nearby ongoing galaxy mergers host exceptionally rich star cluster populations with cluster masses exceeding $10^7~\msun$ \citep{schweizer82,holtzman92,miller97,schweizer98,bastian06}, which are formed due to the perturbation of the interstellar medium (ISM) \citep{schweizer87,ashman92}. The multitude of star clusters suggests that they are useful tracers of past galaxy mergers, especially because they are easily observed up to distances of several tens of megaparsecs. The observed clusters ($>10^4~\msun$) are distributed according to a power law with index $-2$ down to the detection limit \citep{zhang99}. These clusters are thought to be just the `tip of the iceberg', since the initial cluster mass function (ICMF) appears to continue beyond the detection limit and down to {a certain} physical lower mass limit \citep[see e.g.][]{portegieszwart10}.

However, high gas densities and tidal shocks, both of which are prevalent in coalescing galaxies, are known to have a disruptive effect on star clusters \citep{spitzer58,weinberg94b,gieles06}. The destruction rate of star clusters decreases with increasing cluster mass and density \citep{spitzer87,lamers05}.\footnote{Unless the environment in which they reside is so disruptive that it can efficiently destroy a cluster regardless of its mass. In that case, the recently argued scenario in which cluster disruption is mass-independent \citep{whitmore07} can arise \citep{elmegreen10b,kruijssen11}. This would then not be universal, but depends on the environmental conditions.} This indicates that the effects of star cluster disruption could be masked by observational selection effects and go unnoticed in observations, i.e. the brightest and therefore most massive clusters are easiest to detect but also least affected by disruption.

If the ICMF is universal, i.e. all stellar clusters are formed according to a power law with index $-2$ throughout space and time \citep[e.g.][]{kruijssen11d}, then the important role of cluster disruption is supported by the old (`globular') star cluster systems that are observed in nearby spiral and giant elliptical galaxies, which are strongly lacking low-mass clusters with respect to the young populations in presently merging galaxies \citep{vesperini01,fall01,elmegreen10}. For a power law ICMF, the size-of-sample effect would also require that the most massive clusters are formed in the largest bursts of star formation, implying that globular clusters originate from starburst environments. The question thus arises whether or not the disruption of star clusters dominates over their formation in starburst galaxies. This is not easily determined on analytical grounds.

Globular cluster systems are present over most of the galaxy mass range \citep[e.g.][]{peng08}, and as such it is evident that they were not only formed in interactions between massive spiral galaxies; the presence of globular clusters in dwarf galaxies suggests that these also endured starbursts during their early evolution. While the globular clusters of dwarf galaxies are generally metal-poor, the colour distribution of globular clusters is often bimodal in massive spiral galaxies and giant ellipticals \citep{searle78,forbes97,kundu01,peng06}. This colour bimodality may translate into a metallicity bimodality, although it has recently been suggested that it is a relic of a non-linear relation between colour and metallicity \citep{yoon06,chies11,yoon11}. {Regardless of whether the metallicity distribution is bimodal}, a popular explanation for the broad range in metallicities \citep{muratov10} is that the metal-poor clusters preferentially originate from accreted dwarf galaxies \citep{prieto08}, while the metal-rich population was mainly formed in-situ, either by disc instabilities \citep{shapiro10} or in galaxy mergers \citep{ashman92}.

The formation of globular clusters has been investigated in several theoretical and numerical studies \citep{harris94,elmegreen97,bekki02,li04,bournaud08}. As expected from {power law statistics}, these studies all point to dense, gas-rich environments, which are typically correlated with high star formation rate densities. However, the present-day population of globular clusters is not recovered in these studies, because they only concern cluster formation and contain either no description for the further evolution of clusters or a very simplified one. {Separate studies, both analytical and numerical, have shown that} the evolution of (globular) clusters is related to the galactic environment \citep{spitzer87,baumgardt03,lamers05a,gieles06,elmegreen10b,kruijssen11}. To obtain a more complete, quantitative understanding of the origin of present-day globular clusters, it is necessary to consider their formation and further evolution simultaneously.

At present, the most commonly used method to model the evolution of star clusters is through $N$-body simulations \citep{vesperini97b,portegieszwart98,baumgardt03,gieles08,praagman10,renaud11}. However, this method is computationally too expensive to follow the formation and evolution of the entire cluster population. \citet{kruijssen11} therefore introduced a method in which numerical simulations of galaxies are supplemented with a semi-analytic model for the formation and evolution of star clusters, of which the results are consistent with (observed and simulated) formation and destruction rates from the literature. {This model enables us to track the formation and evolution of the entire star cluster population throughout the assembly histories of galaxies.}

As a first effort to understand the (im)balance between the formation and destruction of star clusters in starburst environments, we use the method from \citet{kruijssen11} to model the star cluster populations of galaxy mergers. This allows us to quantify the net effect of a galaxy merger on its cluster population. With this setup, we aim to investigate:
\begin{itemize}
\item[(1)] {the relative importance of} cluster formation and destruction in interacting galaxies;
\item[(2)] whether galaxy mergers can produce the progenitors of present-day metal-rich globular clusters.
\end{itemize}
In Sects.~\ref{sec:clform} and~\ref{sec:clevo} we summarise our model, while the initial conditions of the simulations are presented in Sect.~\ref{sec:init}. The evolution of the star cluster population in galaxy mergers is assessed in Sect.~\ref{sec:evo}, where we also address their sensitivity to model parameters. We end this paper with a summary of our conclusions.

\section{Summary of the model} \label{sec:model}
We model the formation and evolution of star clusters semi-analytically, coupled to a numerical simulation code for galaxy evolution \citep[{{\sc stars}},][]{pelupessy05}. Here we provide a summary of the model, which {was presented and validated} by \citet{kruijssen11}.

\subsection{Galaxy evolution and star cluster formation} \label{sec:clform}
The evolution of the stellar and dark matter components are governed by pure collisionless Newtonian dynamics, calculated using the Barnes-Hut tree method \citep{barnes86}. The particles sample the underlying phase space distribution of positions and velocities and are smoothed on length-scales of approximately 0.2~kpc to maintain the collisionless dynamics and to reduce the noise in the tidal field (which is used for the cluster evolution, see Sect.~\ref{sec:clevo}). The Euler equations for the gas dynamics are solved using smoothed particle hydrodynamics, a Galilean invariant Langrangian method for hydrodynamics based on a particle representation of the fluid \citep{monaghan92}, in the conservative formulation of \citet{springel02}. This is supplemented with a model for the thermodynamic evolution of the gas in order to represent the physics of the interstellar medium (ISM). Photo-electric heating by UV radiation from young stars is included (assuming optically thin transport of non-ionizing photons). The UV field is calculated from stellar UV luminosities derived from stellar population synthesis models \citep{bruzual03}. Line cooling from eight elements (the main constituents of the ISM H and He as well as the elements C, N, O, Ne, Si and Fe) is included. We calculate ionization equilibrium including cosmic ray ionization. Further details of the ISM model can be found in \citet{pelupessy04} and \citet{pelupessy05}.

Star formation is implemented by using a gravitational instability criterion based on the local Jeans mass $M_{\rm J}$:
\begin{equation}
\label{eq:jeanscrit}
M_{\rm J} \equiv \frac{\pi \rho}{6} \left( \frac{\pi s^2}{G \rho} \right)^{3/2} < M_{\rm ref} ,
\end{equation}
where $\rho$ is the local density, $s$ the local sound speed, $G$ the gravitational constant and $M_{\rm ref}$ a reference mass-scale (chosen to correspond to observed giant molecular clouds). This selects cold, dense regions for star formation, which then form stars on a time-scale $\tau_{\rm sf}$ that is set to scale with the local free fall time $t_{\rm ff}$: 
\begin{equation}
\label{eq:tausf}
\tau_{\rm sf}=f_{\rm sf} t_{\rm ff}= \frac{f_{\rm sf}}{\sqrt{4 \pi G \rho}} ,
\end{equation}
where the delay factor $f_{\rm sf} \approx 10$. Numerically, the code stochastically spawns stellar particles from gas particles that are unstable according to Eq.~\ref{eq:jeanscrit} with a probability $1-\exp{(-{\rm d}t/\tau_{\rm sf})}$. The code also assigns a formation time for use by the stellar evolution library, and sets the initial stellar and cluster population mass distributions (see below). Mechanical heating of the interstellar medium by stellar winds from young stars and supernovae is implemented by means of pressure particles \citep{pelupessy04,pelupessy05}, which ensures the strength of feedback is insensitive to numerical resolution effects. In this way, the global efficiency of star formation is determined by the number of young stars needed to quench star formation by UV and supernova heating, which is set by the cooling properties of the gas and the energy input from the stars.

For the purpose of this paper, in which the formation rate of star clusters is compared to their destruction rate, it is essential that we obtain reliable estimates of the star formation rate (SFR). Our model for star formation reproduces the Kennicutt-Schmidt \citep{schmidt59,kennicutt89} pattern of star formation \citep{gerritsen97,pelupessy05} and also gives realistic representation of the relation between molecular ${\rm H}_2$ and star formation \citep{pelupessy06,pelupessy09}. On the other hand our model does simplify the star formation process considerably and this should be kept in mind. First, the absolute scaling of the star formation rates is somewhat uncertain and depends on the choice of parameters. It is mainly sensitive to effective feedback strength, but the feedback strength parameter has been independently constrained within a factor of two by considering the power spectra of the resulting HI distribution maps \citep{pelupessy04,pelupessy05}. Secondly, the Jeans mass argument we use for our star formation model suffers from some limitations. Apart from the fact we take a single reference cloud mass $M_{\rm ref}$ (in principle a more sophisticated model using a cloud spectrum could be constructed), it is also a strictly local criterion: this means that a given point in our simulation is either star forming or not regardless of the immediate environment. A more realistic star formation criterion would try to identify GMC-like structures in the gas distribution and then convert these into stars using its bulk properties (like mass, radius, irradiation and angular momentum) -- possibly on the basis of more detailed modelling results of single GMC calculations. Lastly, our star formation model is based on the presence of a two phase interstellar medium -- and while comparison to structural properties of actual galaxies gives good results \citep{pelupessy04,pelupessy06,pelupessy09} our use of smoothed-particle hydrodynamics (SPH) has known limitations in the representation of strong shocks and instabilities such as the Kevin-Helmholtz instability \citep{agertz07}, which are important in two phase media. This could be checked in future work with shock resolving adaptive mesh refinement (AMR) or moving mesh methods \citep[such as {\sc arepo},][]{springel10}.

Whenever a new star particle is spawned, a `sub-grid' set of star clusters is generated. Their masses are drawn from a power law ICMF with an exponential truncation \citep{schechter76}:
\begin{equation}
\label{eq:ICMF}
N{\rm d}M\propto M^{-2}\exp{(-M/M_\star)}{\rm d}M ,
\end{equation}
where $N$ is the number of clusters, $M$ is the cluster mass, and $M_\star=2.5\times10^6~\msun$ is the exponential truncation mass, which is consistent with high-mass end of the present day mass distribution of globular clusters \citep{fall01,kruijssen09b}. This reflects the observed mass distribution of young star clusters \citep{zhang99,lada03,larsen09,portegieszwart10} and is likely also the ICMF of the majority of globular clusters \citep{kruijssen11d}. We adopt a minimum cluster mass of $M_{\rm min}=10^2~\msun$. The cluster formation rate is assumed to be proportional to the SFR by adopting a constant cluster formation efficiency (CFE) of 90\%, {which is chosen to minimise Poisson noise}. Because it is taken to be constant, the precise value of the CFE is irrelevant and acts as a normalisation of the number of clusters. The remaining 10\% of the mass is considered to be formed as unbound associations or field stars. In the simulation, the field stars are not physically separated from the star clusters, as each star particle contains both clusters and field stars. Because our cluster model is sub-grid, we presently cannot include clusters more massive than about $10^{5.9}~\msun$, which corresponds to the adopted particle mass (see Sect.~\ref{sec:init}). The number of particles in the simulation was chosen to cover the cluster mass range of interest, while ensuring sufficient numerical resolution.

\subsection{Star cluster disruption} \label{sec:clevo}
The further evolution of the stellar clusters is computed with the {\sc space} cluster models \citep{kruijssen08,kruijssen09c}, which include a semi-analytical description of the evolution of the cluster mass and its stellar content. {\sc space} includes stellar evolution from the Padova isochrones \citep{marigo08}, stellar remnant production, remnant kick velocities, dynamical disruption and the evolution of the stellar mass function within the cluster due to the stellar mass dependence of the escape rate. {The cluster evolution model} has been coupled to properties of the tidal field by \citet{kruijssen11} to include tidal evaporation and heating by tidal shocks.

After their formation, the mass evolution of individual clusters is governed by mass loss due to stellar evolution and dynamical disruption:
\begin{equation}
\label{eq:dmdt}
\left(\frac{{\rm d}M}{{\rm d}t}\right) = \left(\frac{{\rm d}M}{{\rm d}t}\right)_{\rm se} + \left(\frac{{\rm d}M}{{\rm d}t}\right)_{\rm dis} ,
\end{equation}
with $M$ the cluster mass and the subscripts `se' and `dis' denoting stellar evolution and disruption, respectively. The mass loss due to stellar evolution is obtained by taking the decrease of the maximum stellar mass over one time step from the Padova models \citep{marigo08}, and integrating the mass function within the cluster over the corresponding mass interval. Upon the removal of these massive stars, the masses of their stellar remnants are added to the cluster mass. The dynamical mass loss is caused by two simultaneous mechanisms. Firstly, the stars in the cluster are driven over the tidal boundary due to two-body relaxation \citep{spitzer87}. Secondly, stars can gain energy from tidal shocks, i.e. fluctuations of the tidal field caused by passages through dense galactic regions such as giant molecular clouds (GMCs) or spiral arms \citep{gieles06,gieles07}.

We parametrize the mass loss due to disruption as
\begin{equation}
\label{eq:dmdtdis}
\left(\frac{{\rm d}M}{{\rm d}t}\right)_{\rm dis}=\left(\frac{{\rm d}M}{{\rm d}t}\right)_{\rm rlx}+\left(\frac{{\rm d}M}{{\rm d}t}\right)_{\rm sh}=-\frac{M}{t_{\rm dis}^{\rm rlx}}-\frac{M}{t_{\rm dis}^{\rm sh}} ,
\end{equation}
where `rlx' and `sh' denote two-body relaxation and tidal shocks, $t_{\rm dis}^{\rm rlx}$ represents the time-scale for disruption by two-body relaxation, and $t_{\rm dis}^{\rm sh}$ the time-scale for disruption by tidal shocks.  Both time-scales are related to the tidal field. The derivation is given in \citet{kruijssen11}, but here we give the final expressions. For $t_{\rm dis}^{\rm rlx}$ the expression is:
\begin{equation}
\label{eq:tevap}
t_{\rm dis}^{\rm rlx}=1.7~{\rm Gyr}~M_4^\gamma\left(\frac{T}{10^{4}~{\rm Gyr}^{-2}}\right)^{-1/2} ,
\end{equation}
where $M_4$ is the cluster mass in units of $10^4~\msun$, $\gamma=0.62$ is the mass dependence of the disruption time-scale \citep{lamers05}, which has a weak dependence on the density profile of the cluster \citep{lamers10}, and $T$ is the tidal field strength. The tidal field strength is taken to be the largest eigenvalue of the tidal tensor.\footnote{By doing so, we ignore potential second-order effects due to the other eigenvalues and the time evolution of the direction of the largest eigenvector \citep{tanikawa10,renaud11}. This choice is made because the erratic tides in galaxy mergers with a gas component obstruct a straightforward implementation of these effects. None the less, the influence on our results should be minor for two reasons. Firstly, our model gives good agreement with direct $N$-body simulations of cluster evolution \citep{baumgardt03}. Secondly, the vast majority of cluster disruption is due to tidal shocks instead of the steady tidal field (see \citealt{kruijssen11} and Sect.~\ref{sec:cfrvsdis}).}, which is determined by numerical differentiation of the force field. Gravity is smoothed on a length-scale of 0.2~kpc and the differentiation interval is 1~per cent of the smoothing length, which ensures that the influence of discreteness noise on cluster disruption is negligible. This was illustrated in \citet{kruijssen11}, where we also showed that the disruption of clusters due to tidal evaporation and tidal shocks is unaffected by passages of single particles for our choice of smoothing length and particle mass. Instead, their disruption is governed by the tidal influence of structures that are well-resolved with our resolution (also see Sect.~\ref{sec:init}). The resulting small influence of numerical resolution on our results is verified in Sect.~\ref{sec:sens}.

For the disruption time-scale due to tidal shocks, the approaches of \citet{gieles07} and \citet{prieto08} can be combined to obtain \citep{kruijssen11}:
\begin{equation}
\label{eq:tsh}
t_{\rm dis}^{\rm sh}=3.1~{\rm Gyr}~M_4\left(\frac{r_{\rm h}}{\rm pc}\right)^{-3}\left(\frac{I_{\rm tid}}{10^{4}~{\rm Gyr}^{-2}}\right)^{-1}\left(\frac{\Delta t}{{\rm Myr}}\right) ,
\end{equation}
where $r_{\rm h}$ is the half-mass radius, $I_{\rm tid}$ is the tidal heating parameter \citep[see][]{gnedin99b,prieto08,kruijssen11}, which follows from the integration of the tidal field over the duration of a shock, and $\Delta t$ the time since the last shock. It reflects the time-scale on which the cluster is heated and is determined individually for each component of the tidal tensor by identifying local minima with sufficient (1$\sigma$) contrast with respect to the preceding maximum \citep[see][]{kruijssen11}. Because the disruption time-scale due to tidal shocks depends on cluster density, it is important to include a description for the half-mass radius. It was recently shown by \citet{gieles11b} that cluster radii pass through two evolutionary phases. Initially, a cluster expands to fill its tidal boundary, during which time the half-mass relaxation time remains constant, i.e. $r_{\rm h}\propto M^{-1/3}$. After filling its tidal boundary, the cluster continues in the `mass-loss dominated regime' along tracks of $r_{\rm h}\propto M^x$, with $x=1/6$ to $1/3$ depending on the escape criterion. The duration of the first phase depends on the initial conditions of cluster formation, while the second phase lasts until the total disruption of the cluster. Since our models have been tuned to agree with the $N$-body simulations of cluster disruption by \citet{baumgardt03}, we assume an evolution of the half-mass radius $r_{\rm h}=4.35~{\rm pc}~(M/10^4~\msun)^{0.225}$, which is consistent with their work \citep[see][]{kruijssen11}. This relation lies in the second evolutionary phase from \citet{gieles11b}, because the clusters from \citet{baumgardt03} are initially filling their tidal boundaries. Because the initial conditions of cluster formation and their impact on the mass-radius relation are quite uncertain, we validate our results using other mass-radius relations in Sect.~\ref{sec:sens}.

Both Eqs.~\ref{eq:tevap} and~\ref{eq:tsh} are calibrated for clusters with a King parameter of $W_0=5$. For other density profiles, the constants in the equations change, but the lifetimes of the clusters are similar. They have been compared and calibrated to the $N$-body simulations of star cluster disruption by \citet{baumgardt03} to ensure their accuracy \citep{kruijssen11}. While most of their simulations concern clusters on circular orbits, we have used those simulations of clusters on eccentric orbits to verify our models for changing tidal fields.

Like the cluster formation rate, it is essential for the purpose of this paper that the estimated disruption rates are reliable. In \citet{kruijssen11}, we have therefore tested our model for a range of different cosmic settings including several isolated disc galaxies and different kinds of galaxy mergers, and compared the results to observations. The simulated age distributions of star clusters in disc galaxies and their correlation with galactocentric radius are in accordance with observational results (see \citealt{bastian11} for an analysis of the cluster population of M83). We also found that in isolated disc galaxies with 15--30\% of their baryonic mass in gas, typically 85\% of the cluster disruption is accounted for by tidal shocks (Eq.~\ref{eq:tsh}), while the remainder is covered by two-body relaxation (Eq.~\ref{eq:tevap}). This is in excellent agreement with a study by \citet{lamers06a}, {who found that in the solar neighbourhood about 80\% of the disruption is contributed by tidal shocks}. The high relative contribution of tidal shocks to cluster disruption shows that the tidal field in an isolated galaxy is far from smooth due to encounters with GMCs and spiral arms. This is an important similarity to galaxy mergers, in which the tidal field also varies, albeit to a larger extent. We have also applied our models to the Antennae galaxies (Kruijssen \& Bastian, in prep.) and find good agreement with the observed cluster age and mass distributions from \citet{whitmore07}. These results provide a good starting point to apply our model to galaxy mergers and follow the formation and evolution of the entire star cluster population for different galactic histories.
\begin{table*}\centering
\caption[]{\label{tab:discs}
     \sf Details of the initial conditions for the disc galaxy models.}
\begin{tabular}{c c c c c c c c c c c c}
\hline
${\rm ID}$ & $f_{\rm gas}$ & ${M_{\rm vir}}^a$ & $z$ & $\lambda$ & $N_{\rm halo}$ & $N_{\rm gas}$ & $N_{\rm disc}^{\rm star}$ & $N_{\rm bulge}^{\rm star}$ & ${M_{\rm part}^{\rm halo}}^a$ & ${M_{\rm part}^{\rm bary}}^a$ & ${\rm Comments}$
\\\hline
${\rm 1dA}$    &   $   0.20$     &   $  10^{12}$     & $  2 $  &    $ 0.05$     &     $10^6$     &   $  10250$     & $    41000 $   &   $   10000 $  & $10^6$ & $8\times10^5$ & ${\rm     low~gas~fraction }$     \\
${\rm 1dB}$    &   $   0.30$     &   $  10^{12}$     & $  2 $  &    $ 0.05$     &     $10^6$     &   $  15375$     & $    35875 $   &   $   10000 $  & $10^6$ & $8\times10^5$ & ${\rm  standard~model}$ \\
${\rm 1dC}$    &   $   0.50$     &   $  10^{12}$     & $  2 $  &    $ 0.05$     &     $10^6$     &   $  25625$     & $    25625 $   &   $   10000 $  & $10^6$ & $8\times10^5$ & ${\rm     high~gas~fraction  }$    \\
${\rm 1dD}$    &   $   0.30$     &   $  5\times10^{11}$ &   $2$   &    $   0.05$   &       $5\times10^5$   &        $  7688     $   &   $    17938$  &       $ 5000$ & $10^6$ & $8\times10^5$ &  ${\rm   half~mass }$\\
${\rm 1dE}$    &   $   0.30$     &   $  10^{12}$     & $  2 $  &    $ 0.05$     &     $10^6$     &   $  15375$     & $    35875 $     & $    0         $  & $10^6$ & $8\times10^5$ & ${\rm     no~bulge  }$         \\
${\rm 1dF}$    &   $   0.30$     &   $  10^{11}$     & $  2 $  &    $ 0.05$     &     $10^6$     &   $  15375$     & $    35875 $   &   $   10000 $  & $10^5$ & $8\times10^4$ & ${\rm     low~mass }$          \\
${\rm 1dG}$    &   $   0.30$     &   $  10^{12}$     & $  2 $  &    $ 0.10$     &     $10^6$     &   $  15375$     & $    35875 $   &   $   10000 $  & $10^6$ & $8\times10^5$ & ${\rm     high~spin    }$       \\
${\rm 1dH}$    &   $   0.30$     &   $  10^{12}$     & $  0 $  &    $ 0.05$     &     $10^6$     &   $  15375$     & $    35875 $   &   $   10000 $  & $10^6$ & $8\times10^5$ & ${\rm     low~concentration}$        \\
${\rm 1dI  }$  &     $ 0.30  $   &     $10^{12}$     &   $5   $&     $0.05  $   &     $10^6$     &     $15375  $   &   $  35875   $ &     $ 10000   $& $10^6$ & $8\times10^5$ &   ${\rm   high~concentration}$        \\
\hline
\multicolumn{12}{l}{\sf $^a$In solar masses ($\msun$).}
\end{tabular} 
\end{table*}

\subsection{Initial conditions} \label{sec:init}
We use the set of simulations described in \citet{kruijssen11} and summarise the adopted parameter sets here. The simulations follow the evolution of the star cluster population in isolated disc galaxies and galaxy mergers. The disc galaxies are generated with parameters that can be related to the outcomes of cosmological $\Lambda$CDM galaxy formation models \citep{mo98,springel05b}. They consist of a dark halo with a \citet{hernquist90} profile, an exponential stellar disc, a stellar bulge (except for one model) and a thin gaseous disc, constructed to be in self gravitating equilibrium if evolved autonomously \citep{springel05b}. The dark matter haloes have concentration parameters related to their total masses and condensation redshifts according to \citet{bullock01}, {implying that for a fixed mass the halo concentration increases with redshift}. The total mass is related to the virial velocity $V_{\rm vir}$ and the Hubble constant $H(z)$ at redshift $z$ as $M_{\rm vir}=V_{\rm vir}^3/[10GH(z)]$. For all galaxies, the baryonic disc is constituted by a gaseous and stellar component, having a mass fraction $m_{\rm d}=0.041$ of the total mass, while the bulge (when included) consists of a stellar component only, having a mass fraction $m_{\rm b}=0.008$ of the total mass. The fraction of total angular momentum that is constituted by the disc ($j_{\rm d}$) is taken identical to $m_{\rm d}$. The scale-length of the bulge and the vertical scale-length of the disc are 0.2 times the radial scale-length of the disc, which is determined by the {degree of rotation \citep{mo98} through the spin parameter $\lambda\equiv J |E| / GM_{\rm vir}^{5/2}$, in which $J$ is the angular momentum of the halo and $E$ its total energy}. We have chosen the parameter sets to cover a reasonable spread in galaxy properties, specifically the gas fraction of the baryonic disc $f_{\rm gas}$, their total mass  $M_{\rm vir}$, the spin parameter $\lambda$ and the presence of a bulge. The resulting disc galaxy model parameters can be found in Table~\ref{tab:discs}, which lists $f_{\rm gas}$, $M_{\rm vir}$, $\lambda$, the number of particles in the different components of the model galaxies, and the particle masses of the halo particles $M_{\rm part}^{\rm halo}$ and baryonic particles $M_{\rm part}^{\rm bary}$. {For each set of initial conditions, $M_{\rm part}^{\rm halo}$ and $M_{\rm part}^{\rm bary}$ are chosen to be very similar. Our particle resolution is a trade-off between enabling the formation of high-mass star clusters (which are limited by the particle mass due to their sub-grid treatment, see Sect.~\ref{sec:clform}) and resolving the galaxy dynamics.} We verified in \citet{kruijssen11} that the adopted resolution is sufficient to reliably model the cluster disruption, because the influence of encounters with single particles is negligible compared to the tidal perturbation of clusters by more massive, resolved structures. 

\begin{table}\centering
\caption[]{\label{tab:mergers}
     \sf Details of the initial conditions for the galaxy merger models.}
\begin{tabular}{c c r r r r r c}
\hline
${\rm ID}$ & ${\rm Discs}$ & $\theta_1$ & $\phi_1$ & $\theta_2$ & $\phi_2$ & ${R_{\rm p}}^a$ & ${\rm Type}^b$
\\\hline
${\rm 1m1}$      &    ${\rm AA}$    &      $0$         & $    0$     &   $   0   $      &   $  0  $     &    $   6   $   &   ${\rm   PP }$     \\
${\rm 1m2}$      &    ${\rm BB}$    &      $0$         & $    0$     &   $   0   $      &   $  0  $     &    $   6   $   &   ${\rm   PP }$       \\
${\rm 1m3}$      &    ${\rm CC}$    &     $0$        &  $    0$    &    $   0   $     &    $ 0   $     &    $  6    $  &    ${\rm  PP  }$    \\
${\rm 1m4}$      &    ${\rm BD}$    &      $0$         & $    0$     &   $   0   $      &   $  0  $     &    $   6   $   &   ${\rm   PP }$         \\
${\rm 1m5}$    &    ${\rm EE}$     &      $0$        &    $ 0  $   &   $   0  $       &   $  0   $     &  $    6 $     &    ${\rm  PP  }$         \\
${\rm 1m6}$    &    ${\rm FF}$     &       $0$        &    $ 0  $   &   $   0  $       &   $  0   $     &  $    6 $     &    ${\rm  PP  }$         \\
${\rm 1m7}$    &    ${\rm GG}$     &     $0$        &    $ 0  $   &   $   0  $       &   $  0   $     &  $    6 $     &    ${\rm  PP  }$         \\
${\rm 1m8}$    &    ${\rm HH}$     &     $0$        &    $ 0  $   &   $   0  $       &   $  0   $     &  $    6 $     &    ${\rm  PP  }$         \\
${\rm 1m9}$    &    ${\rm II}$         &      $0$        &    $ 0     $&    $  0$         &     $0        $&     $ 6     $ &     ${\rm PP  }$         \\\hline
${\rm 1m10}$      &    ${\rm BB}$    &      $60$       &$     45$   &  $   -45   $    &$    -30$    & $     6 $     & ${\rm     PP}_{\rm i}$           \\
${\rm 1m11}$      &    ${\rm BB}$    &      $180$     &$     0 $    &  $    0      $   & $    0    $    & $     6 $     & ${\rm     PR}$           \\
${\rm 1m12}$      &    ${\rm BB}$    &      $-120$   &$     45$   &  $    -45  $    &$     -30$    & $     6 $     & ${\rm     PR}_{\rm i}$           \\
${\rm 1m13}$      &    ${\rm BB}$    &      $180$     & $    0 $    &  $    180 $    & $    0    $   &   $   6  $    &  ${\rm    RR }$         \\
${\rm 1m14}$      &    ${\rm BB}$    &      $-120$   &  $   45$   &$      135$     & $    -30 $   &   $   6  $    &  ${\rm    RR}_{\rm i}$           \\
${\rm 1m15}$    &    ${\rm BB}$     &      $0$        &    $ 0    $ &  $    0       $  &    $ 0       $ &     $ 12  $  &    ${\rm  PP}_{\rm w}$           \\
${\rm 1m16}$    &    ${\rm CG}$     &      $-120$  &   $  45 $  & $     -45  $   &    $ -30   $ &     $ 10  $  &    ${\rm  PR}_{\rm i}$           \\\hline
${\rm 1m17}$      &    ${\rm BB}$    &      $0$       &     $0$   &    $ 71       $&  $  30   $ &     $ 6  $    &           ${\rm Barnes     }$      \\
${\rm 1m18}$      &    ${\rm BB}$    &      $-109$       &$     90$   &   $  71   $    &   $ 90  $ &     $ 6    $  &    ${\rm  Barnes    }$       \\
${\rm 1m19}$      &    ${\rm BB}$    &      $-109$       &$     -3$0   &  $   71  $     &  $  -30$    &  $    6$      &${\rm      Barnes}$           \\
${\rm 1m20}$      &    ${\rm BB}$    &      $-109$       &$     30$   &   $  180 $      & $   0  $  &    $  6   $   &   ${\rm   Barnes   }$        \\
${\rm 1m21}$      &    ${\rm BB}$    &      $0      $ &     $0$   &  $   71  $     &$    90  $  &    $  6  $    &           ${\rm Barnes     }$      \\
${\rm 1m22}$      &    ${\rm BB}$    &      $-109$       &$     -30 $  &    $ 71  $     &  $  30  $  &  $    6 $     & ${\rm     Barnes }$          \\
${\rm 1m23}$      &    ${\rm BB}$    &      $-109$       &$     30  $ &     $71   $    &   $ -30  $  &  $    6 $     & ${\rm     Barnes }$          \\
${\rm 1m24}$      &    ${\rm BB}$    &      $-109$       &$     90  $ &     $180 $      & $   0   $ &   $   6  $    &   ${\rm   Barnes   }$        \\\hline
\multicolumn{8}{l}{\sf $^a$In kpc. All angles are in degrees.}\\
\multicolumn{8}{l}{\sf $^b$Indicates prograde-prograde (PP), prograde-retrograde (PR),}\\
\multicolumn{8}{l}{\sf retrograde-retrograde (RR) or `Barnes' (see text). Subscripts `i' and}\\
\multicolumn{8}{l}{\sf `w' denote inclined/near-polar and wide orbits, respectively.}
\end{tabular} 
\end{table}
In the galaxy merger simulations, the disc galaxies follow Keplerian parabolic orbital trajectories with initial separations of approximately 200~kpc. The actual orbit will decay due to dynamical friction, which leads to the merging of the galaxies. The orbital geometry of an interaction is characterised by the directions of the angular momentum vectors of the two galaxy discs and the pericentre distance of the parabolic orbit $R_{\rm peri}$. The angular momentum vectors of the galaxies are determined in spherical coordinates by angles $\theta$ (rotation perpendicular to the orbital plane) and $\phi$ (rotation in the orbital plane). These and other relevant parameters are listed in Table~\ref{tab:mergers}, which covers three subsets of simulations. The first set of eight runs follow a common configuration, in which the discs rotate in the orbital plane. They are used to test the influence of the properties of the progenitor discs. The six subsequent runs are aimed at investigating the impact of orbital parameters on the star cluster population. We rotate the progenitor discs to include retrograde rotation and near-polar orbits, which represent the opposite extreme with respect to the co-planar configurations of the first eight runs. A wider orbit and a `random' major merger are also considered. The third group contains the eight `random' configurations from \citet{hopkins09} \citep[see][]{barnes88}, which together sample the phase space of possible orbital geometries.
\begin{figure*}
\center
\resizebox{17.5cm}{!}{\includegraphics{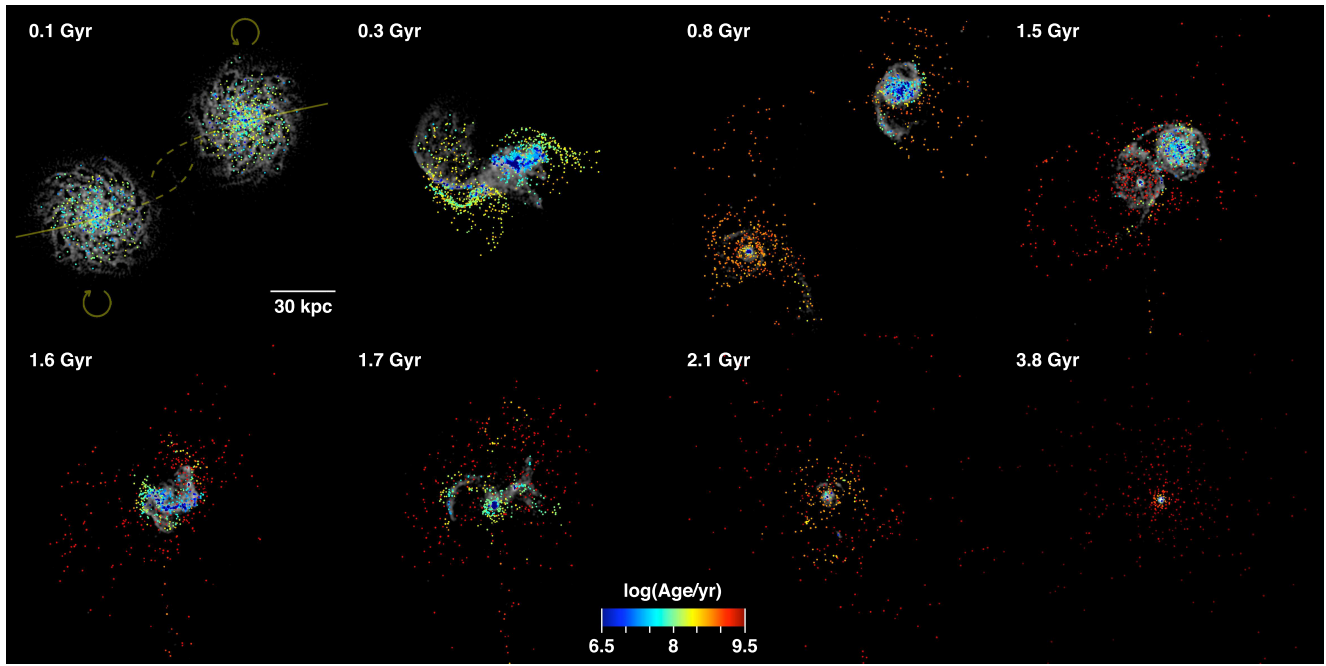}}
\caption[]{\label{fig:snapshots}\sf
       Evolution of the star cluster population during a galaxy merger (simulation {\tt 1m11} from Table~\ref{tab:mergers}). The surface density of the gas is displayed in greyscale, while the particles that contain star clusters are shown in colours denoting the ages of the clusters as indicated by the legend. The subsequent images show the collision at { eight} characteristic moments: $t=0.1$~Gyr, briefly before the first passage; { $t=0.3$~Gyr, just after the first passage;} $t=0.8$~Gyr, in between the first and second passage; { $t=1.5$~Gyr, just before the second passage;} $t=1.6$~Gyr, just after the second passage; { $t=1.7$~Gyr just before the final coalescence; $t=2.1$~Gyr, during the coalescence and the infall of remaining gas clouds;} $t=3.8$~Gyr, when only a merger remnant is left. See {\tt http://www.mpa-garching.mpg.de/\~{}diederik/1m11clusters.html} for a movie of the full time sequence.
                 }
\end{figure*}

All galaxies are generated without any star clusters, and we set $t=0$ after 300~Myr of evolution to initialise the cluster population. As described in Sect.~\ref{sec:model}, the clusters have masses between 10$^2$ and $\sim10^{5.9}~\msun$, following a \citet{schechter76} type initial mass function. The chemical composition of the clusters is set to solar metallicity,\footnote{This choice only affects the stellar evolutionary mass loss.} and we assume a King parameter of $W_0=5$.

\section{Evolution of the star cluster population} \label{sec:evo}
In this section, we apply our model to the evolution of the star cluster population in galaxy mergers. We start out by explaining an illustrative example, before discussing the other simulations and trends with merger properties.

\subsection{Illustrative example} \label{sec:example}
Figure~\ref{fig:snapshots} shows a classical sequence of the evolution of a galaxy merger simulation together with the results from our cluster evolution model (simulation {\tt 1m11} from Table~\ref{tab:mergers}). The panels in Fig.~\ref{fig:snapshots} show the distributions of gas and star clusters at different times during the interaction. The first image displays the galaxies as they approach each other for their first passage (at $t=0.1$~Gyr), when the tidal interaction between the galaxies is still relatively weak and the SFR is at a low-to-intermediate level ($\sim 6~\msun~{\rm yr}^{-1}$). The spatial distribution of star clusters is restricted to both galaxy discs, where the gas resides from which they are formed, and their destruction rate is still low since it is only driven by the internal galactic tidal field and encounters between clusters and GMCs.

{ In the second and third images of Fig.~\ref{fig:snapshots} ($t=0.3$--0.8~Gyr), the galaxies are shown (briefly) after their first passage.} By this time, the gravitational interaction has produced extended tidal tails. Most star clusters still follow the morphology of the gas because they have just been formed in a large starburst (about $50~\msun$~yr$^{-1}$) that was triggered by the angular momentum loss and consequent inflow of the gas during the first pericentre passage. { In the second image, the total number of clusters reaches a peak, with an increase of $\sim 40\%$ with respect to the first panel, but in the third image only one third of this number is left.}\footnote{{The total number of clusters depends on the lower mass limit (see Sects.~\ref{sec:cfrvsdis} and~\ref{sec:fsurv}), which in our simulations is taken to be $10^2~\msun$.}} This is caused by the large central gas density that drives the starburst, prompting a stronger increase of the tidal perturbation of star clusters than of their formation rate. Some intermediate age clusters have been ejected from the discs by the interaction. They represent the first star cluster constituents of a stellar halo forming around the two galaxies. The mechanisms of {\it cluster migration} and {\it natural selection} that were identified in \citet{kruijssen11} are evident here: clusters are escaping the dense regions in which they were formed and those clusters in quiescent, low-density parts of the galaxies have higher chances of survival. As a result, the mean disruption rate decreases with cluster age. The enhanced disruption of young clusters due to their tidal interaction with the primordial environment was named the {\it cruel cradle effect} in \citet{kruijssen11b}, and can affect clusters up to ages of $\tau \sim 200$~Myr, contrary to the early disruption of clusters by gas expulsion (`infant mortality'), which takes place on a gas expulsion time-scale of $\sim 10$~Myr \citep{lada03,goodwin06,pelupessy11}.

As the galaxies proceed to merge, the effects of the interaction intensify. The { fifth and sixth panels} of Fig.~\ref{fig:snapshots} display the galaxies during the short interval between the second passage and their final coalescence ($t=1.6$--1.7~Gyr), in a configuration that is similar to the `Antennae' galaxies \citep[NGC 4038/9, see][]{karl10}. During this phase, the remaining gas is funnelled towards the centres of the galaxies, where it cools to form large numbers of stars and star clusters. This second starburst is accompanied by an even stronger increase of the cluster destruction rate, this time disrupting well over 50\% of all clusters. A large number of clusters is ejected from the central region into the stellar halo that surrounds the galaxies. Away from the turmoil, these clusters will be able to survive the coalescence of the galaxies. By the time the merger is completed, most of the surviving clusters will have formed at this moment or around the time of the first snapshot in Fig.~\ref{fig:snapshots} \citep{kruijssen11}.

When the merger is completed, as is shown in the last image of Fig.~\ref{fig:snapshots} ($t=3.8$~Gyr), the system has transformed into a giant elliptical galaxy, in which the star cluster system has dispersed into the stellar halo. The formation rate of clusters drops to a minimum after the merger, due to the depletion of the gas during the starbursts. The gas depletion also affects the disruption rate: it causes the typical tidal shock strength to decrease after the coalescence of the two galaxies, implying that massive, dense clusters are no longer affected by disruption. However, 
the migration of clusters towards radial orbits during violent relaxation causes the tidal shock heating to remain high, as clusters on very eccentric orbits are being disrupted by bulge shocks. At this stage, low-mass star clusters are preferentially disrupted, and their destruction leads to an increase of the mean cluster mass.

\subsection{Formation versus destruction} \label{sec:cfrvsdis}
\begin{figure}
\center\resizebox{8cm}{!}{\includegraphics{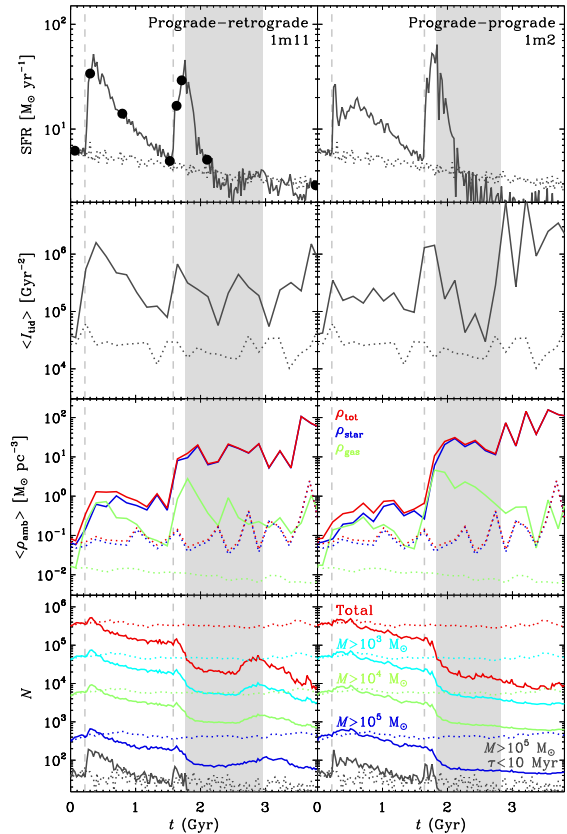}}
\caption[]{\label{fig:sfhncx}\sf
       Evolution of the star formation rate (SFR, first row), mean tidal heating ($\langle I_{\rm tid}\rangle$, second row), mean ambient gas, stellar and total densities ($\langle \rho_{\rm amb}\rangle$, third row), and the number of star clusters ($N$, fourth row) for two different galaxy merger simulations. The left-hand panels show the results for the prograde-retrograde encounter from Fig.~\ref{fig:snapshots} (simulation {\tt 1m11} from Table~\ref{tab:mergers}), while the right-hand panels represent a similar encounter with both galaxies rotating in the prograde direction (anticlockwise in the configuration of Fig.~\ref{fig:snapshots}, simulation {\tt 1m2} from Table~\ref{tab:mergers}). The thick dots mark the moments that are displayed in Fig.~\ref{fig:snapshots}. The number of star clusters in the bottom panels is shown for different cluster mass cuts ($\log{(M/\msun)}>\{2,3,4,5\}$). {The bottom lines also include an age limit ($\tau<10$~Myr), and show the number of young massive clusters as a function of time.} The dotted curves denote the results for the two disc galaxies evolving in isolation. The vertical dashed lines indicate the times of first and second pericentre passage and the shaded areas specify the time interval over which the final coalescence occurs.
                 }
\end{figure}
We have tested the generality of these results by analysing the full set of simulations from Sect.~\ref{sec:init}. The results of two simulations are shown in Fig.~\ref{fig:sfhncx}, where the star formation history (SFH) as well as the time-evolution of the mean tidal shock heating, the mean ambient densities of gas and stars,\footnote{These are determined by using the approach from \citet{casertano85}, where the density is averaged over a sphere with radius equal to the distance to the $N$th nearest neighbour. We adopt $N=7$.} and the number of star clusters for different cluster mass cuts are shown. The figure also includes a comparison with the two disc galaxies evolving in isolation. Just after the pericentre passages, the galaxies exhibit a pronounced increase of the SFR (0.5--1~dex), {but an even stronger increase of the mean tidal shock heating (1--1.5~dex), leading to a decrease of the total number of clusters by nearly two orders of magnitude towards the end of the simulations}. It is not the tidal field strength itself,\footnote{In fact, the mean tidal field strength is lower in the merger simulations than in the isolated disc galaxies. This is caused by several factors, but is mainly related to the rapid destruction of clusters by tidal shocks, which works most efficiently in the regions of a merger where the absolute tidal field strength is also higher. The resulting cluster population is biased towards larger galactocentric radii than in the isolated disc case, implying that the mean tidal field strength is lower. Being the dominant source of the cluster destruction, the mean tidal shock heating is not affected by this selection effect (see Fig.~\ref{fig:sfhncx}).} but the frequency and strength of tidal shocks which leads to the enhanced disruption of clusters.

If the (environmentally dependent) increase of the tidal disruption is neglected, the total number of clusters increases by a factor of 5--6 during the first pericentre passage for simulation {\tt 1m11} compared to the progenitor discs evolving in isolation. For both galaxy mergers in Fig.~\ref{fig:sfhncx}, the destruction of clusters is most prominent after the second pericentre passage and during the coalescence of the galaxies. As indicated in Sect.~\ref{sec:example}, the decrease of the number of clusters is largest for the lowest cluster masses, which is clearly seen in the number evolution for different cluster mass cuts in Fig.~\ref{fig:sfhncx}. After the mergers are completed, their SFRs become lower than would have been the case had the galaxies evolved in isolation. At this stage, the mean tidal shock heating remains relatively high due to the disruption of clusters on radial orbits (see below).

The densities in Fig.~\ref{fig:sfhncx} illustrate that both the starbursts and the episodic increase of the tidal shock heating are caused by the growth of the ambient gas density. The stellar density is almost always higher than the gas density, but during the pericentre passages and final coalescence, the development of peaks in $\langle I_{\rm tid}\rangle$ correlates with $\langle \rho_{\rm gas}\rangle$ rather than $\langle \rho_{\rm star}\rangle$: contrary to $\langle \rho_{\rm star}\rangle$, which remains at a constant, high level after each increase, $\langle \rho_{\rm gas}\rangle$ and $\langle I_{\rm tid}\rangle$ return to lower values. The reason that the tidal shock heating is dominated by the ambient gas density instead of the (higher) stellar density is that the gas is more structured, which produces faster tidal shocks that cannot be absorbed by the adiabatic expansion of the clusters. Only after the final coalescence of the galaxies, the tidal shock heating is dominated by the stellar density. This occurs after $t=3$~Gyr, when most of the cluster disruption is caused by the tidal shocking of clusters on highly eccentric orbits, which are falling in from the tidal tails or have migrated to a radial orbit due to violent relaxation.

\begin{figure}
\center\resizebox{8cm}{!}{\includegraphics{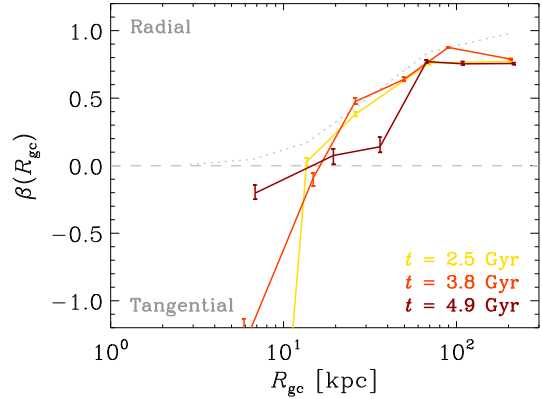}}\\
\caption[]{\label{fig:beta}\sf
      Orbital anisotropy of the surviving star clusters $\beta$ (see Eq.~\ref{eq:beta}) as a function of galactocentric radius. The relation is shown at different times (indicated by the legend) during the final coalescence and in the merger remnant of simulation {\tt 1m11}. The clusters are binned using an equal numbers of clusters per bin, with the error bars denoting the error on the mean. Preferentially radial and tangential orbits are separated by the horizontal dashed line at $\beta=0$, which indicates orbital isotropy. The dotted line is included for reference and shows the radial dependence of $\beta$ for the parametrization of the velocity ellipsoid from \citet{aguilar88} with anisotropy radius $R_{\rm A}=30$~kpc. There are indications that the globular cluster system of M87 has a similar anisotropy profile \citep{strader11}.
                 }
\end{figure}
In time, the secular evolution of a merger remnant decreases the orbital anisotropy of the surviving clusters through the disruption of clusters on eccentric orbits. To quantify the orbital anisotropy of the star cluster system as a function of galactocentric radius $R_{\rm gc}$, the anisotropy parameter is defined as
\begin{equation}
\label{eq:beta}
\beta(R_{\rm gc})=1-\frac{\langle v_{\rm t}^2\rangle}{2\langle v_{\rm r}^2\rangle} ,
\end{equation}
where $\langle v_{\rm r}^2\rangle$ is the mean square radial velocity in a radial bin centered at $R_{\rm gc}$, and $\langle v_{\rm t}^2\rangle$ is the mean square tangential velocity. For the isotropic case we have $\beta=0$, while $\beta>0$ and $\beta<0$ indicate preferentially radial and tangential orbits, respectively. In Fig.~\ref{fig:beta}, the time evolution of the anisotropy parameter is shown as a function of galactocentric radius, for snapshots during and after the final coalescence in simulation {\tt 1m11}. Up to $t=3.8$~Gyr, the kinematics of the inner $\sim 10$~kpc are dominated by rotation because the encounter is co-planar. Outside this radius, the cluster system quickly becomes radially anisotropic, with an anisotropy radius close to $R_{\rm A}=30$~kpc. As mentioned earlier, this is caused by the infall of clusters from the tidal tails and the migration to eccentric orbits due to violent relaxation. However, after $t=3.8$~Gyr the radial anisotropy disappears due to the destruction of clusters on radial orbits (also see Fig.~\ref{fig:sfhncx}). In the inner $\sim 10$~kpc, the cluster system also evolves towards isotropy. The anisotropy radius increases to $R_{\rm A}=50$--60~kpc, similar to the result found for globular clusters from accreted dwarf galaxies \citep{prieto08}. This shows that it is not straightforward to distinguish between in-situ and ex-situ formation based on the orbital (an)isotropy of a cluster system.

The results in Fig.~\ref{fig:sfhncx} suggest that galaxy mergers efficiently disrupt star clusters, in apparent contradiction with the rich star cluster populations that are observed in colliding galaxies such as the Antennae \citep{whitmore99} and M51 \citep{bastian05}. However, our above analysis concerns the entire star cluster population in a merger, while observations are naturally constrained to bright clusters, which are typically massive and young. When limiting our results to the young ($\la10$~Myr) and massive ($\ga10^4$--$10^5~\msun$) clusters that are easily detected in observations \citep[e.g.][]{zhang99}, Fig.~\ref{fig:sfhncx} shows that the number of clusters that would be `observed' from our simulation temporarily increases by more than a factor of three during starbursts. {The increase in observed galaxy mergers may be even higher than this \citep[see e.g.][]{schweizer96,miller97,zepf99}, but the factor of three increase is a lower limit for a number of reasons. Firstly, simulation {\tt 1m11} is one of the more monotonously disruptive merger simulations in our sample (second from the right in Fig.~\ref{fig:survssfr}, see below). It can be contrasted with simulation {\tt 1m7}, which exhibits the highest degree of variation over the course of the merger: the number of young massive clusters is temporarily boosted by a factor of 6-15 during the time interval $t=0.5$--$2.2$~Gyr, whereas the total number of clusters eventually settles at only 7\% of the amount the progenitor galaxies would have had in isolation. Secondly, we did not include a variable CFE, which may increase with the star formation rate density (\citealt{goddard10,adamo11}, although see \citealt{silvavilla11}). This could imply that the number of clusters increases by an additional factor of 2--3 during starbursts. Thirdly, our cluster masses are limited to $10^{5.9}~\msun$, which obstructs the formation of the extremely massive ($\geq10^6~\msun$) clusters that are observed in galaxy mergers and for which the relative increase with respect to quiescent galaxies is most evident. These effects could conspire to yield a transient relative increase of young massive clusters during the starbursts of up to a factor of $\sim30$.} None the less, in terms of numbers, a star cluster population is dominated by the unseen low-mass star clusters that are effectively destroyed during a merger before they reach ages much older than a few tens of Myr.

\subsection{A generalised relation for star cluster survival} \label{sec:fsurv}
For all simulations, the results are in accordance with those shown in Fig.~\ref{fig:sfhncx}, as they exhibit a very similar increase of the mean tidal shock heating and corresponding decrease of the number of clusters during the merger. The number of clusters in our merger remnants is always 2--50\% of the amount that the two discs would have contained in isolation. Much of the variation is caused by the different orbital geometries. Retrograde, co-planar encounters lead to enhanced angular momentum loss of the gas and correspondingly stronger starbursts and greater destruction of clusters, decreasing their number by up to a factor of 50. Galaxies on wide or inclined orbits such that they follow near-polar trajectories prompt a weaker effect due to a less pronounced gas inflow, yielding a decrease of about a factor of 2--10. 

\begin{figure}
\center\resizebox{8cm}{!}{\includegraphics{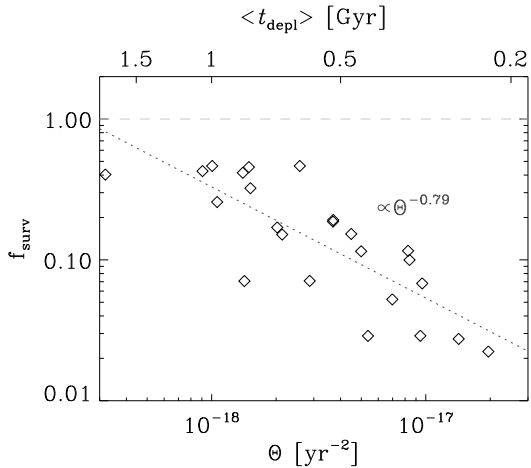}}\\
\caption[]{\label{fig:survssfr}\sf
      Star cluster `survival fraction' $f_{\rm surv}$ as a function of the starburst intensity parameter $\Theta$ (see text and Eq.~\ref{eq:fsurv}). The corresponding logarithmic mean of the gas depletion time-scale $\langle t_{\rm depl}\rangle\equiv\Theta^{-1/2}$ of the two starbursts during a merger is indicated along the top axis. Symbols denote the 24 merger simulations from Table~\ref{tab:mergers}, and the dotted line gives a power law fit.
                       }
\end{figure}
We find that the total number of surviving clusters strongly decreases with increasing peak SFR. This trend is a consequence of the disruptive power of dense, star-forming environments. To quantify this trend across all simulations, one can define the ratio of the number of clusters in the galaxy mergers relative to the number of clusters in the isolated progenitor galaxies ($f_{\rm surv}$). We have tested the dependence of $f_{\rm surv}$ on several generalised forms of the peak SFR, by normalising ${\rm SFR}_{\rm peak}$ to the galaxy stellar, gas or baryonic mass ($M_{\rm star}$, $M_{\rm gas}$ or $M_{\rm bary}$, respectively). It is found that $f_{\rm surv}$ most tightly correlates with ${\rm SFR}_{\rm peak}/M_{\rm gas}$. The simulated galaxy mergers typically experience two starbursts (during the pericentre passages), and therefore the value of $f_{\rm surv}$ in a merger remnant includes the effect of two starbursts. For the case of a single starburst, we write a power law formulation $f_{\rm surv}=C[{\rm SFR}_{\rm peak}/(M_{\rm gas}~{\rm yr}^{-1})]^\alpha$. For the number of clusters in a merger remnant this implies:
\begin{equation}
\label{eq:fsurv}
 f_{\rm surv}=f_{\rm surv,1}f_{\rm surv,2}=C^2\left(\frac{{\rm SFR}_{\rm peak,1}}{M_{\rm gas,1}~{\rm yr}^{-1}}\frac{{\rm SFR}_{\rm peak,2}}{M_{\rm gas,2}~{\rm yr}^{-1}}\right)^\alpha\equiv C^2\Theta^\alpha ,
 \end{equation}
where subscripts 1 and 2 indicate the first and second starbursts, respectively, and we have defined a starburst intensity parameter $\Theta\equiv{\rm SFR}_{\rm peak,1}{\rm SFR}_{\rm peak,2}/(M_{\rm gas,1}M_{\rm gas,2}~{\rm yr}^{-2})$. 

Figure~\ref{fig:survssfr} shows $f_{\rm surv}(\Theta)$ when measuring $f_{\rm surv}$ at $t=4.8$~Gyr, which is typically 2~Gyr after the completion of each merger. The `survival fraction' $f_{\rm surv}$ very clearly decreases with increasing starburst intensity $\Theta$. A simple power law fit to the data points in Fig.~\ref{fig:survssfr} gives $C=4.5\pm1.5\times10^8$ and $\alpha=0.79\pm0.13$ over almost two orders of magnitude in $\Theta$. While the deviation of some data points is as high as 0.5~dex, the correlation is quite remarkable considering the wide range of boundary conditions that is covered (see Tables~\ref{tab:discs} and~\ref{tab:mergers}). The relation flattens when increasing the lower mass limit of the clusters, since massive clusters are less rapidly disrupted than low-mass clusters. For our models, this can be approximated to reasonable accuracy by $C=4.5\times10^{-8}M_{\rm min,2}^{2}$ and $\alpha=-0.77+0.22\log{M_{\rm min,2}}$, for $M_{\rm min,2}\equiv M_{\rm min}/10^2~\msun$ and $10^2\leq M_{\rm min}/\msun\leq 10^4$. At larger minimum masses, $C$ and $\alpha$ remain constant, although it is uncertain to what extent this may be the result of our maximum mass limit of $10^{5.9}~\msun$. Another source of uncertainty is the variation of the ICMF truncation mass $M_\star$ with the galactic environment or SFR \citep{bastian08,larsen09,kruijssen11d}. If $M_\star$ increases with the SFR, galaxy mergers naturally yield a net production of star clusters with masses larger than the value of $M_\star$ in quiescent progenitor galaxies. This has been reported to be about $M_\star\sim2\times10^5~\msun$ \citep{larsen09}, which thus indicates the mass scale that separates net cluster destruction at low cluster masses from a net production at higher masses.

If we write Eq.~\ref{eq:fsurv} in terms of the gas depletion time-scale $t_{\rm depl}\equiv M_{\rm gas}/{\rm SFR}_{\rm peak}$, the above results in a generalised expression for $f_{\rm surv}$ after a single starburst, which is given by
\begin{equation}
\label{eq:tdepl}
 f_{\rm surv}(M>M_{\rm min})=4.5\times10^{-8}M_{\rm min,2}^{2}\left(\frac{t_{\rm depl}}{{\rm yr}}\right)^{0.77-0.22\log{M_{\rm min,2}}} ,
\end{equation}
for $10^2\leq M_{\rm min}/\msun\leq 10^4$ and $0.1\leq t_{\rm depl}/{\rm Gyr} \leq 3$. A naive extrapolation of this expression gives a net increase of the number of star clusters during starbursts above $\sim 3\times10^5~\msun$, very similar to the approximate value of $M_\star$ in quiescent galaxies. Since we neglect any environmental variation of $M_\star$, this similarity is a coincidence that potentially allows Eq.~\ref{eq:tdepl} to be extended to $M_{\rm min}>10^4~\msun$. This will need to be addressed in a future work that does account for a variation of the truncation mass. Because $t_{\rm depl}^{-1}$ is a measure of the intensity of the starburst, Eq.~\ref{eq:tdepl} reflects that clusters are disrupted by the dense star-forming environment. Figure~\ref{fig:survssfr} has thus shown that, ironically, star formation kills.

The enhanced cluster disruption rate during starbursts is also demonstrated by the displacement of the peaks in the cluster age distribution and SFH that was presented in \citet[Figs.~15 and~17]{kruijssen11}. Star clusters that are formed during starbursts are found to experience such an elevated disruption rate that they have severely lower survival chances than clusters formed in quiescent environments. As a result, the peaks in the cluster age distribution and SFH can differ by up to 200~Myr, with the bulk of the surviving star clusters being formed prior to the height of star formation \citep[also see][]{chien10}. This implies that the cluster age distribution may indicate the occurrence of a starburst, but (depending on the strength of the starburst) cannot always be used to accurately determine its time or duration. For quiescent galaxies, the age distribution of star clusters does reflect the SFH quite well, modulo a correction for cluster disruption \citep[cf.][]{lamers05,bastian11b}.

\subsection{The cluster mass function} \label{sec:cmf}
\begin{figure}
\center\resizebox{8cm}{!}{\includegraphics{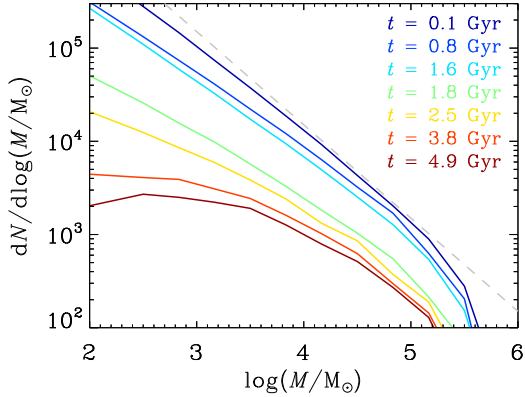}}
\caption[]{\label{fig:histgc}\sf
      Evolution of the mass distribution of star clusters during merger simulation {\tt 1m11}. Shown are the distributions at different times $t$. { We refer to Fig.~\ref{fig:snapshots} for the merger episodes to which these times correspond.} As time progresses, the distribution shifts downwards due to the net destruction of clusters. The slope of the initial mass distribution is shown as a dashed line, which would have closely resembled the mass distribution at all times had the two galaxies evolved in isolation.
                 }
\end{figure}
The preferential destruction of the low-mass clusters causes the initially scale-free (except for the Schechter-type truncation) cluster mass distribution to develop a characteristic mass,\footnote{This requires that tidal shocks most efficiently disrupt low-mass clusters, i.e. that the density of clusters increases with their mass, and therefore would not occur for mass-radius relations $r_{\rm h}\propto M^\delta$ with $\delta\geq1/3$. However, such a strong correlation is not supported by observational evidence \citep[e.g.][]{harris96,larsen04b,bastian05}. We explore the dependence of the results on the mass-radius relation in Sect.~\ref{sec:sens}.} which is shown in Fig.~\ref{fig:histgc} for simulation {\tt 1m11}. Most of this transformation occurs during the final coalescence of the galaxies from $t=1.8$~Gyr onwards, when the disruption rate is no longer high enough to affect the most massive clusters, but is still sufficient to efficiently destroy low-mass clusters. At $t=4.9$~Gyr, the peak mass is about $10^{2.7}~\msun$ and increases steadily. Tentative evidence for this peaked form of the cluster mass distribution is also found in observations of recent merger remnants \citep[e.g.][]{goudfrooij04,goudfrooij07}, {albeit at higher masses (we refer the reader to Sect.~\ref{sec:sens} for a discussion of the variation of the modelled peak mass with model parameters).}

The surviving population of clusters that were formed before and during a merger bears hints of {\it observed} globular cluster systems. First of all, the spatial configuration of these clusters is comparable to that of the globular cluster population of the Milky Way \citep{harris96}, giant elliptical galaxies \citep{harris09b} and young merger remnants \citep{schweizer96}, following a power law density profile with index $-3.2$ in the outer parts, which is the approximate behaviour of a de Vaucouleurs profile \citep{vaucouleurs48}. Secondly, the development of a peak in the mass distribution at a mass of $10^{2.7}~\msun$ is very suggestive. This peak mass is still lower than the characteristic mass of globular cluster systems \citep[$10^5~\msun$,][]{harris96}, which would be attained during the several billions of years of star cluster disruption following a high-redshift merger until the present day \citep{vesperini01,fall01,kruijssen09b}, possibly also due to subsequent collisions with other galaxies. Lastly, due to the high peak SFR, a merger can produce a population of clusters that extends to higher masses than for isolated galaxies, in agreement with observations \citep{bastian08,kruijssen11d}. It is therefore capable of producing clusters with the initial masses needed to survive for a Hubble time.

\begin{figure}
\center\resizebox{8cm}{!}{\includegraphics{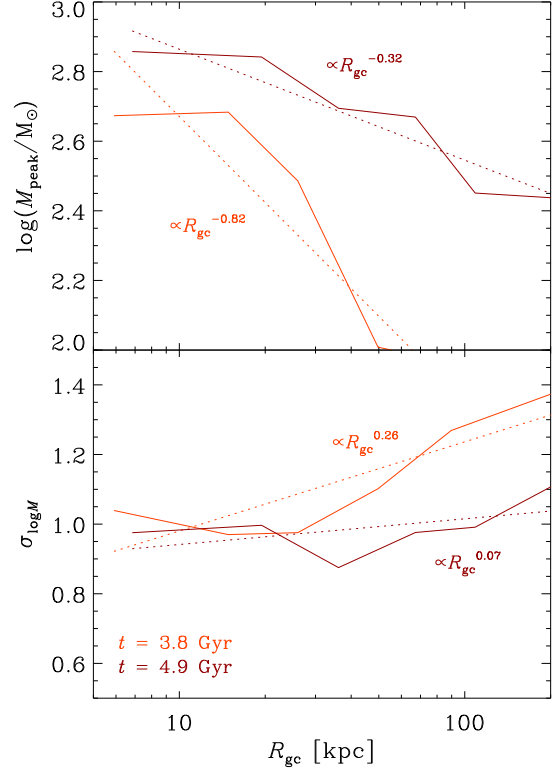}}\\
\caption[]{\label{fig:gcmfrad}\sf
      Radial variation of the peak mass (top panel) and dispersion (bottom panel) of the cluster mass distribution at two different times in a merger remnant. Solid lines show the data from simulation {\tt 1m11} in bins with an equal number of clusters per bin, while dotted lines represent power law fits with slopes as indicated by the labels.
                 }
\end{figure}

Shortly after the completion of a merger, secular cluster disruption increases the characteristic mass by 0.3--0.4~dex per Gyr for the next two gigayears (also see Fig.~\ref{fig:gcmfrad}). If the merger took place in the early universe ($\ga 9$~Gyr ago), the characteristic mass would thus have the time to evolve to that of observed globular cluster systems. The combination of several globular cluster-like characteristics (the spatial distribution, characteristic mass, and high maximum cluster mass) would not be reproduced without the starburst and gas depletion, {the migration of clusters into the halo,} and the enhanced disruption occurring during the starburst.

It is tempting to interpret the existence of a peak in the mass distribution of the surviving star clusters as the early formation of a globular cluster system. Such a scenario was first proposed by \citet{ashman92}. If this were the case, the orbital kinematics of the clusters should evolve towards a state in which there is no radial trend of the characteristic mass, as is the case for the globular cluster systems of the Milky Way \citep{harris96} and M87 \citep{vesperini03}. This could proceed by orbital migration or by the destruction of clusters with certain orbital characteristics. The violent relaxation occurring during galaxy mergers is indeed efficient at ejecting clusters from their original environment \citep{prieto08,bastian09,kruijssen11}, which is also shown by the assembly of the stellar halo { between the second and sixth} images of Fig.~\ref{fig:snapshots}. However, this is not sufficient to prevent a radial trend of the peak mass, because during and after the final coalescence the disruption rate in the galaxy centre is still higher than in the outskirts. We have fitted log-normal functions to the cluster mass distribution as a function of galactocentric radius to quantify its radial variation. Together with the dispersion of the cluster mass distribution, this is shown in Fig.~\ref{fig:gcmfrad} for simulation {\tt 1m11} at two different times after the completion of the merger. The peak of the mass distribution of the surviving clusters in our simulations is initially ($t=3.8$~Gyr) constant for radii $R_{\rm gc}<20$~kpc, but at larger radii it depends on the galactocentric radius as $M_{\rm peak}\propto R_{\rm gc}^{-0.8}$, signifying a higher disruption rate in the galaxy centre. By $t=4.9$~Gyr, the destruction of clusters on radially anisotropic orbits (see Fig.~\ref{fig:beta}) has led to a shallower dependence of $M_{\rm peak}\propto R_{\rm gc}^{-0.3}$. The dispersion of the mass distribution follows a similar evolution. It is almost fully insensitive to $R_{\rm gc}$ at $t=4.9$~Gyr and steadily decreases with time.

While the radial variation of $M_{\rm peak}$ evolves towards the radially independent form that characterises observed globular cluster systems, there is no guarantee that this will still be the case at the present day. If the further evolution of the merger remnant is quiescent, a radial trend of the characteristic mass might be reintroduced after a few Gyr, since the disruption of clusters proceeds more rapidly near the galaxy centre than at large radii. It may be possible to erase this radial dependence again later on, for instance due to perturbations of the cluster orbits by minor mergers \citep[cf.][]{qu11}, or any other perturbations that make the galaxy potential deviate from spherical symmetry \citep[see also][]{fall01}. The details of the further evolution of the cluster mass distribution and its spatial variation will depend on the cosmic environment, and cannot be followed in a major merger simulation. Whichever cosmic conditions may govern the further evolution of the cluster population, the relative universality \citep[see e.g.][]{jordan07} of the globular cluster luminosity function suggests that these conditions should be commonplace if the cluster populations of our merger remnants are indeed the progenitors of present-day globular cluster systems. This is discussed further in Sect.~\ref{sec:concl}.

\begin{figure}
\center\resizebox{8cm}{!}{\includegraphics{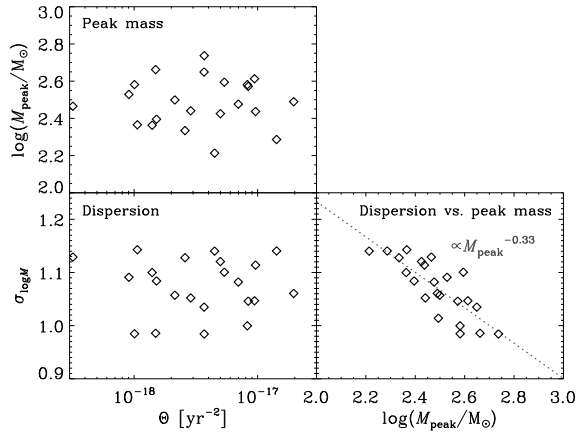}}
\caption[]{\label{fig:gcmfssfr}\sf
      Peak mass $M_{\rm peak}$ and logarithmic dispersion $\sigma_{\log{M}}$ as a function of the starburst intensity parameter $\Theta$ (left-hand panels) and of each other (right-hand panel). Symbols denote the 24 merger simulations from Table~\ref{tab:mergers}. The cluster mass functions are taken at $t=4.8$~Gyr, and clusters formed during the last 500~Myr are excluded. In the right-hand panel, the dotted line gives a power law fit, which becomes shallower as the population ages.
                 }
\end{figure}
Contrary to the survival fraction of star clusters (see Fig.~\ref{fig:survssfr} and Eq.~\ref{eq:tdepl}), the peak mass and dispersion of the cluster mass distribution does not exhibit any variation with starburst intensity, which is shown in the left-hand panels of Fig.~\ref{fig:gcmfssfr}. This contrast with the survival fraction arises because $f_{\rm surv}$ is set at a different time during the merger than the shape of the cluster mass distribution. The cluster destruction rate is highest when the gas density peaks, at the height of the interaction and starbursts. The galactic environment at that stage is in fact so disruptive that star clusters can be destroyed irrespective of their masses. The slope of the cluster mass distribution is therefore fairly constant at the times when the majority of all clusters is being destroyed. Only after the onset of galaxy coalescence the frequency and strength of tidal shocks drop to a level at which the most massive clusters are able to survive, implying that low-mass clusters are preferentially disrupted. This leads to a flattening of the cluster mass distribution at the low-mass end. Figure~\ref{fig:histgc} nicely illustrates this, as the slope of the mass distribution mostly changes after 1.8~Gyr. The shape of the mass distribution is thus set during the aftermath of the merger and therefore does not correlate with the cluster survival fraction, which is determined by the starburst intensity at earlier times.

The right-hand panel in Fig.~\ref{fig:gcmfssfr} does show that the dispersion of the cluster mass distribution correlates with the peak mass as $\sigma_{\log{M}}\propto M_{\rm peak}^{-0.33}$. This relation arises because the high-mass end of the mass distribution does not vary much across all simulations, implying that low-mass cluster disruption sets both the peak mass and the width of the distribution. Present-day globular cluster systems have a peak mass of about $\log{(M_{\rm peak}/\msun)}=5$--5.3 and a dispersion of $\sigma_{\log{M}}=0.4$--0.5 \citep{jordan07}. Extrapolation of the power law fit in Fig.~\ref{fig:gcmfssfr} to such peak masses yields too small dispersions ($\sim0.2$), indicating that the relation between dispersion and peak mass should flatten as cluster disruption proceeds. There are hints of a dependence of $\sigma_{\log{M}}$ on $M_{\rm peak}$ for old globular cluster populations in the Virgo Cluster \citep[e.g.][]{jordan07}, but indeed the correlation is not as strong as we find for the cluster populations of young merger remnants. A direct comparison between the dispersion in our simulations and in the observed systems is obstructed by the variation of the truncation mass $M_\star$ among galaxies in the Virgo Cluster, which is not included in our models.

\subsection{Sensitivity of results to model assumptions} \label{sec:sens}
For the simulations that are used in this paper, we have adopted certain initial conditions and made a number of assumptions. In this section, the presented results are tested for any dependences on two key assumptions. We verify the influence of (1) the evolution of cluster radii and (2) the adopted particle resolution. The former is important because the dominant cluster destruction mechanism (tidal shock disruption) scales with cluster density, while the latter should be checked to see if our simulations indeed resolve the gas structure that disrupts clusters (see the discussion in Sect.~\ref{sec:init}).

\begin{figure}
\center\resizebox{8cm}{!}{\includegraphics{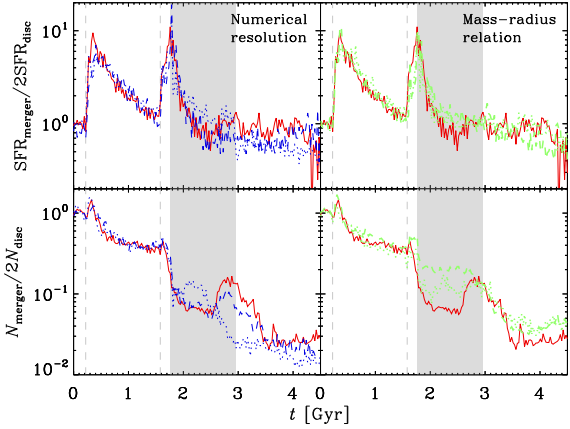}}
\caption[]{\label{fig:sens}\sf
      Influence of numerical resolution (left-hand panels) and the star cluster mass--radius relation (right-hand panels) on the SFR (top panels) and the number of clusters (bottom panels) in merger simulation {\tt 1m11}. The SFR and number of clusters are normalised to twice the value they have in each corresponding isolated disc simulation. The red {solid} lines show the benchmark result from Fig.~\ref{fig:sfhncx} with a mass--radius relation and number of particles as described in Sects.~\ref{sec:clevo} and~\ref{sec:init}. In the left-hand panels, the dotted blue lines indicate the same simulation as {\tt 1m11} but with double the particle resolution, whereas the dashed blue lines denote the same simulation with half the particle resolution. In the right-hand panels, the dotted green lines represent the same simulation as {\tt 1m11} but using a constant cluster radius $r_{\rm h}=3.75$~pc independent of cluster mass, while the dashed green lines mark the same simulation with a constant cluster density $r_{\rm h}/{\rm pc}=3.75(M/10^4~\msun)^{1/3}$.
                 }
\end{figure}
Figure~\ref{fig:sens} shows the effect of the numerical resolution and the star cluster mass-radius relation on the SFH and the time evolution of the total number of clusters over the course of our merger simulation {\tt 1m11}. {The numerical resolution is changed by a factor of two up and down, while the added mass-radius relations imply a constant radius $r_{\rm h}=3.75$~pc and constant density $r_{\rm h}/{\rm pc}=3.75(M/10^4~\msun)^{1/3}$.} The SFH and the number of clusters are both normalised to their disc values to enable a straightforward comparison. The figure demonstrates that the relative change of the SFR and number of star clusters in the galaxy merger simulations with respect to isolated discs is not much affected by numerical resolution and the cluster mass-radius relation. At the end of the simulations, all differences are of the order of the statistical scatter. The only significant deviation seems to occur between $t=2$~Gyr~and $t=3.5$~Gyr, when the number of clusters in the reference simulation is relatively unstable with respect to the other simulations. This can be traced to the stochastic variation of the SFR in the top panels (and thus the cluster formation rate). The top-right panel illustrates the statistical spread of the SFH over different realisations of the model, because the cluster mass-radius relation does not influence the SFH. In other words, the three shown SFHs are the result of identical boundary conditions. It implies that the dip in the SFH of the reference simulation at $t\sim2.4$~Gyr and the subsequent rise of the SFR are stochastic. Due to the ongoing disruption of star clusters, this statistical variation is magnified in their number evolution \citep[bottom panels, also see][]{kruijssen11}. We can therefore conclude from Fig.~\ref{fig:sens} that the time evolution of the number of clusters in galaxy mergers relative to isolated discs is not influenced by numerical resolution and the cluster mass-radius relation.

The absolute number of clusters in the simulations does change for different mass-radius relations, because star cluster disruption in galaxy discs and galaxy mergers is dominated by tidal shocks, for which the cluster disruption time-scale depends on the cluster density. However, when comparing the merger simulations to the corresponding isolated discs, this difference is offset by a similar change of the absolute number of clusters in both cases. The same holds for any possible variation due to numerical resolution \citep[cf.][Fig.~4]{kruijssen11}. This validates the results presented in Sects.~\ref{sec:example}--\ref{sec:fsurv}, where we looked at the impact of galaxy mergers relative to isolated dics. 

\begin{figure}
\center\resizebox{8cm}{!}{\includegraphics{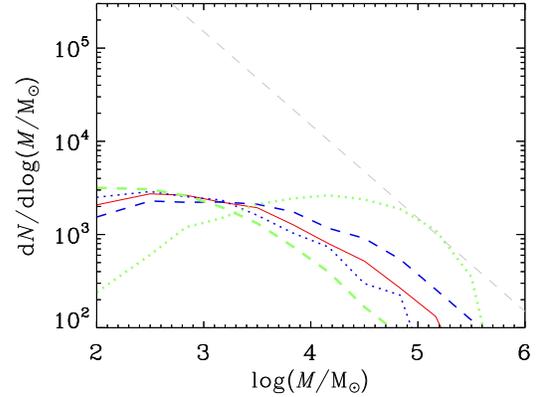}}
\caption[]{\label{fig:gcmfsens}\sf
      The star cluster mass distribution at $t=4.8$~Gyr for varying numerical resolution and mass--radius relations. As in Fig.~\ref{fig:sens}, the red {solid} line indicates the reference simulation {\tt 1m11}, with blue lines denoting double (dotted) and half (dashed) the numerical resolution, and green lines representing constant cluster radii (dotted) and densities (dashed). To account for the statistical scatter in Fig.~\ref{fig:sens}, all distributions are normalised to the number of clusters in simulation {\tt 1m11} and clusters formed during the last 500~Myr are excluded. As in Fig.~\ref{fig:histgc}, the slope of the initial mass distribution is shown as a dashed line.
                 }
\end{figure}
We did not yet verify the impact of the cluster mass-radius relation on the cluster mass distribution (cf. Sect.~\ref{sec:cmf}). The low-mass (disruption-dominated) end of the cluster mass distribution attains a slope equal to the mass dependence of the disruption time-scale \citep{fall01}. This slope and the peak mass should therefore be sensitive to the mass-radius relation (see the expression for the tidal shock disruption time-scale in Eq.~\ref{eq:tsh}). In Fig.~\ref{fig:gcmfsens} we show the cluster mass function in the merger remnant of simulation {\tt 1m11} at $t=4.8$~Gyr, for the same set of simulations as in Fig.~\ref{fig:sens}. Again, the numerical resolution of the simulations does not strongly influence the result. {It only affects the high-mass truncation of the mass distribution because the particle mass limits the cluster mass (see Sect.~\ref{sec:clform}).} However (and as expected), the cluster mass distribution does vary for different mass-radius relations. The question thus arises which range is possibly covered by the normalisation and exponent of the actual mass-radius relation. We included an extensive discussion of this topic in \citet{kruijssen11}, in which we motivated our choice of the mass-radius relation by comparison to $N$-body simulations of dissolving star clusters \citep{baumgardt03,kuepper08}. The adopted mass-radius relation ($r_{\rm h}=r_{\rm h,4}[M/10^4~\msun]^\delta$, with $r_{\rm h,4}=4.35$~pc and $\delta=0.225$) is consistent with the mass-loss dominated regime of \citet{gieles11b}. However, the evolution of star clusters is initially dominated by expansion and the mass-radius relation can not be expressed by a single power law for the entire cluster history. \citet{gieles11b} show that the exponent of the mass-radius relation varies from $\delta=-0.25$ during the expansion-dominated phase to $\delta=0.17$ in the mass-loss dominated phase. This is close to our adopted value, which should thus be taken as an upper limit to the exponent throughout the evolutionary histories of star clusters. This rules out the green dashed line in Fig.~\ref{fig:gcmfsens}, which shows the cluster mass distribution for $\delta=1/3$.

Before estimating lower and upper limits to the peak mass of the cluster mass distribution, the choice of the normalisation of the mass-radius relation $r_{\rm h,4}$ should also be evaluated because it affects the peak mass. The typical radius of young star clusters of $10^4~\msun$ is $r_{\rm h}\sim3.75$~pc \citep{larsen04b}, which we adopted as the normalisation for the constant-radius and constant-density relations in Figs.~\ref{fig:sens} and~\ref{fig:gcmfsens}, whereas clusters in our reference simulation have $r_{\rm h,4}=4.35$~pc to match $N$-body simulations. The mean and median half-mass radii\footnote{These are determined by assuming that light traces mass, i.e. that the clusters are not mass-segregated. This may underestimate the radii by up to a factor of 1.5, potentially depending on cluster mass (see \citealt{gieles11b} for a discussion).} of Galactic globular clusters fall in the same range, with 4.3~pc and 3.0~pc, respectively \citep[updated 2010~version]{harris96}, showing little variation with cluster mass. Such little change after a Hubble time of evolution with respect to young clusters suggests that the adopted normalisations in Fig.~\ref{fig:gcmfsens} are representative of typical star cluster sizes for populations of any age. Combining this with the exclusion of the constant-density ($\delta=1/3$) simulation above, we can conclude that our reference simulation ($\delta=0.225$) provides a {\it lower limit} for the peak mass of $10^{2.5}~\msun$. The green dotted line in Fig.~\ref{fig:gcmfsens} shows the mass distribution for $\delta=0$, while observations of young clusters suggest $\delta=0.1$ \citep{larsen04b}. A peak mass of $10^4~\msun$ can thus be interpreted as an {\it upper limit} to the peak mass in a merger remnant. Our conclusions should be specified further in a future work, by sampling the initial cluster radii from a certain (possibly cluster mass dependent) distribution function and including a more physically motivated description for the further radius evolution.

\section{Conclusions} \label{sec:concl}
We have performed a numerical study of major mergers of comparable-mass disc galaxies, complemented with a sub-grid model for the ongoing formation and evolution of their star cluster populations. The simulations have been used to address the relative contributions of cluster formation and disruption over the course of a galaxy merger, and to investigate the potential formation of the metal-rich part of a globular cluster system. The main results from our model are as follows.

\begin{itemize}
\item[(i)] During a galaxy merger, the total number of star clusters decreases. The increase of the star formation rate during merger-induced starbursts is compensated by a stronger increase of the cluster disruption due to tidal shock heating by dense gas.
\item[(ii)] Although during certain episodes the destruction rate is high enough to disrupt clusters independently of their mass, over the entire course of a merger low-mass clusters are most strongly affected by the destruction. When considering increasingly massive clusters, their number decreases by a smaller amount during a merger. If the cluster sample is limited to massive and young clusters to mimic observational selection effects, the net destruction cannot be detected and changes to a transient increase of the number of clusters during the starbursts, in agreement with observational results.
\item[(iii)] The relative decrease of the number of clusters is stronger for higher peak star formation rates, because the enhanced formation and destruction of clusters are both caused by the high gas density. This trend is weaker for higher masses and may be reversed above $M\sim 1$--$3\times10^5~\msun$, where a stronger starburst may produce {\it more} clusters than a weak starburst. In Eq.~\ref{eq:tdepl}, we provide a generalised expression for the survival fraction of clusters as a function of the gas depletion time-scale, which reflects the intensity of the starburst.
\item[(iv)] {The peaks in the cluster age distribution and star formation history can be offset with respect to each other due to the elevated cluster disruption rate at the height of a starburst. This offset can be as large as 200~Myr \citep{kruijssen11}, which implies that while the cluster age distribution can be used to reveal the occurrence of a starburst, it cannot necessarily be used to determine its time or duration.}
\item[(v)] The orbital kinematics of the star clusters in a merger remnant are isotropic within galactocentric radii of $\sim 50$--60~kpc due to the destruction of clusters on highly eccentric orbits. This value is similar to the result for the accretion of globular clusters from satellite dwarf galaxies \citep{prieto08}, which shows that it may not be possible to distinguish between in-situ and ex-situ cluster formation based on solely the orbital (an)isotropy of the cluster population.
\item[(vi)] The preferential destruction of low-mass clusters causes the power law initial cluster mass function to develop a peak at a mass of about $10^{2.5}~\msun$ during the final coalescence of the galaxies. This is a lower limit, as the precise value depends on the relation between cluster mass and radius, with the post-merger peak mass potentially reaching up to $10^4~\msun$ if the cluster radii are completely unrelated to their masses. The peak mass only weakly correlates with galactocentric radius due to the destruction of clusters on radially anisotropic orbits, and (for the adopted mass-radius relation) increases by about 0.3--0.4~dex per Gyr after the completion of a merger. Young to intermediate-age ($\sim 2$~Gyr old) merger remnants should display a peak in the star cluster mass distribution at about $10^3~\msun$ due to the destruction of low-mass clusters (see Figs.~\ref{fig:histgc} and~\ref{fig:gcmfsens}).
\item[(vii)] After a merger is completed, the star cluster population is similar to what a young globular cluster system would look like. Firstly, the ejection of clusters from star-forming regions into the stellar halo produces a spatial distribution that is comparable to that of globular clusters. Secondly, the peaked cluster mass distribution is intermediate to that of young massive clusters and old globular clusters. Thirdly, the high star formation rate during a merger is capable of producing clusters that are massive enough to survive for a Hubble time.
\end{itemize}

Interestingly, the high disruption rate after (globular) cluster formation could lead to a mass distribution with a peak mass of $10^3$--$10^4~\msun$ on such a short time-scale that only little further disruption is required to obtain the current peak mass of the globular cluster mass distribution. This would imply that even the subset of clusters on the widest orbits around their host galaxies would be able to reach it before the present day. If the ICMF of globular clusters had a Schechter-type truncation at the high-mass end \citep{kruijssen11d}, any further disruption would not yield an additional increase of the peak mass because it then saturates at about 10\% of the truncation mass\footnote{This percentage applies if the ICMF of globular clusters followed a power law with index $-2$ below the truncation, and the mass dependence of the disruption time-scale is $\gamma\sim0.7$ \citep{gieles09,kruijssen09b}.} \citep{gieles09}. This would thus lead to a `universal' globular cluster mass function, independent of galactocentric radius and current galactic environment. The current peak mass of globular cluster systems throughout the universe indeed happens to be $\sim 2\times10^5~\msun$ \citep[e.g.][]{jordan07}, around 10\% lower than the estimated truncation mass of their ICMF \citep{kruijssen09b}.

The increased cluster disruption rate in galaxy mergers is driven by the high gas densities that also cause the burst of star formation. This indicates that the mechanism of enhanced disruption is not necessarily constrained to major mergers, and can be generalised to any environment with a high gas density and a correspondingly high SFR. While major mergers may provide an efficient formation channel for globular cluster populations, they are not a prerequisite. Any extremely high-density, dynamically active, star-forming environment -- be it in a starburst dwarf galaxy, during bulge assembly, in an unstable high-redshift disc or in a major merger -- would cause the enhanced disruption of clusters at young ages. The clusters that eventually survive are characterised by a more quiescent evolution due to cluster migration and natural selection \citep{kruijssen09b,elmegreen10,elmegreen10b,kruijssen11}. Indeed, the wide variety of galaxy types with remarkably similar globular cluster mass distributions is hard to explain if cluster disruption is governed by the present-day environment, and suggests that the bulk of the disruption occurred at the epoch of globular cluster formation, when the host galaxies were likely more similar. A generalisation to all dense environments is supported by dwarf galaxies like Fornax, which has not undergone a major merger and yet harbours a handful of globular clusters \citep{shapley39,hodge61} {that presumably formed in a starburst during the early formation of the galaxy}. If such a generalisation to all dense environments indeed holds, it would suggest that globular cluster populations may be the inevitable outcomes of the large starbursts occurring in the early universe.

\section*{Acknowledgments}
Our calculations were performed at the computing facilities of Leiden Observatory. This research is supported by the Netherlands Advanced School for Astronomy (NOVA), the Leids Kerkhoven-Bosscha Fonds (LKBF) and the Netherlands Organisation for ScientiÞc Research (NWO), grants 021.001.038, 639.073.803, and 643.200.503, as well as by the DFG cluster of excellence `Origin and Structure of the Universe' (www.universe-cluster.de). We thank to Mark Gieles and Oleg Gnedin for stimulating discussions and comments on an early version of the paper. JMDK gratefully acknowledges the hospitality of the Institute of Astronomy in Cambridge, where a large part of this work took place.

\bibliographystyle{mn2e2}
\bibliography{mybib}

\begin{thebibliography}{117}
\expandafter\ifx\csname natexlab\endcsname\relax\def\natexlab#1{#1}\fi
\small
\bibitem[{{Adamo} {et~al.}(2011){Adamo}, {{\"O}stlin}, \&
  {Zackrisson}}]{adamo11}
{Adamo} A., {{\"O}stlin} G., {Zackrisson} E., 2011, \mnras, 417, 1904

\bibitem[{{Agertz} {et~al.}(2007){Agertz}, {Moore}, {Stadel}, {Potter},
  {Miniati}, {Read}, {Mayer}, {Gawryszczak}, {Kravtsov}, {Nordlund}, {Pearce},
  {Quilis}, {Rudd}, {Springel}, {Stone}, {Tasker}, {Teyssier}, {Wadsley}, \&
  {Walder}}]{agertz07}
{Agertz} O., {Moore} B., {Stadel} J., {Potter} D., {Miniati} F., {Read} J.,
  {Mayer} L., {Gawryszczak} A., {Kravtsov} A., {Nordlund} {\AA}., {Pearce} F.,
  {Quilis} V., {Rudd} D., {Springel} V., {Stone} J., {Tasker} E., {Teyssier}
  R., {Wadsley} J., {Walder} R., 2007, \mnras, 380, 963

\bibitem[{{Aguilar} {et~al.}(1988){Aguilar}, {Hut}, \& {Ostriker}}]{aguilar88}
{Aguilar} L., {Hut} P., {Ostriker} J.~P., 1988, \apj, 335, 720

\bibitem[{{Ashman} \& {Zepf}(1992)}]{ashman92}
{Ashman} K.~M., {Zepf} S.~E., 1992, \apj, 384, 50

\bibitem[{{Barnes} \& {Hut}(1986)}]{barnes86}
{Barnes} J., {Hut} P., 1986, \nat, 324, 446

\bibitem[{{Barnes}(1988)}]{barnes88}
{Barnes} J.~E., 1988, \apj, 331, 699

\bibitem[{{Barnes} \& {Hernquist}(1996)}]{barnes96}
{Barnes} J.~E., {Hernquist} L., 1996, \apj, 471, 115

\bibitem[{{Bastian}(2008)}]{bastian08}
{Bastian} N., 2008, \mnras, 390, 759

\bibitem[{{Bastian} {et~al.}(2011{\natexlab{a}}){Bastian}, {Adamo}, {Gieles},
  {Lamers}, {Larsen}, {Silva-Villa}, {Smith}, {Kotulla}, {Konstantopoulos},
  {Trancho}, \& {Zackrisson}}]{bastian11}
{Bastian} N., {Adamo} A., {Gieles} M., {Lamers} H.~J.~G.~L.~M., {Larsen} S.~S.,
  {Silva-Villa} E., {Smith} L.~J., {Kotulla} R., {Konstantopoulos} I.~S.,
  {Trancho} G., {Zackrisson} E., 2011{\natexlab{a}}, \mnras, 417, L6

\bibitem[{{Bastian} {et~al.}(2011{\natexlab{b}}){Bastian}, {Adamo}, {Gieles},
  {Silva-Villa}, {Lamers}, {Larsen}, {Smith}, {Konstantopoulos}, \&
  {Zackrisson}}]{bastian11b}
{Bastian} N., {Adamo} A., {Gieles} M., {Silva-Villa} E., {Lamers}
  H.~J.~G.~L.~M., {Larsen} S.~S., {Smith} L.~J., {Konstantopoulos} I.~S.,
  {Zackrisson} E., 2011{\natexlab{b}}, \mnras~in press, {{\tt ArXiV:1109.6015}}

\bibitem[{{Bastian} {et~al.}(2005){Bastian}, {Gieles}, {Lamers}, {Scheepmaker},
  \& {De Grijs}}]{bastian05}
{Bastian} N., {Gieles} M., {Lamers} H.~J.~G.~L.~M., {Scheepmaker} R.~A., {De
  Grijs} R., 2005, \aap, 431, 905

\bibitem[{{Bastian} {et~al.}(2006){Bastian}, {Saglia}, {Goudfrooij},
  {Kissler-Patig}, {Maraston}, {Schweizer}, \& {Zoccali}}]{bastian06}
{Bastian} N., {Saglia} R.~P., {Goudfrooij} P., {Kissler-Patig} M., {Maraston}
  C., {Schweizer} F., {Zoccali} M., 2006, \aap, 448, 881

\bibitem[{{Bastian} {et~al.}(2009){Bastian}, {Trancho}, {Konstantopoulos}, \&
  {Miller}}]{bastian09}
{Bastian} N., {Trancho} G., {Konstantopoulos} I.~S., {Miller} B.~W., 2009,
  \apj, 701, 607

\bibitem[{{Baumgardt} \& {Makino}(2003)}]{baumgardt03}
{Baumgardt} H., {Makino} J., 2003, \mnras, 340, 227

\bibitem[{{Bekki} {et~al.}(2002){Bekki}, {Forbes}, {Beasley}, \&
  {Couch}}]{bekki02}
{Bekki} K., {Forbes} D.~A., {Beasley} M.~A., {Couch} W.~J., 2002, \mnras, 335,
  1176

\bibitem[{{Bournaud} {et~al.}(2008){Bournaud}, {Duc}, \&
  {Emsellem}}]{bournaud08}
{Bournaud} F., {Duc} P., {Emsellem} E., 2008, \mnras, 389, L8

\bibitem[{{Bressert} {et~al.}(2010){Bressert}, {Bastian}, {Gutermuth},
  {Megeath}, {Allen}, {Evans}, {Rebull}, {Hatchell}, {Johnstone}, {Bourke},
  {Cieza}, {Harvey}, {Merin}, {Ray}, \& {Tothill}}]{bressert10}
{Bressert} E., {Bastian} N., {Gutermuth} R., {Megeath} S.~T., {Allen} L.,
  {Evans} II N.~J., {Rebull} L.~M., {Hatchell} J., {Johnstone} D., {Bourke}
  T.~L., {Cieza} L.~A., {Harvey} P.~M., {Merin} B., {Ray} T.~P., {Tothill}
  N.~F.~H., 2010, \mnras, 409, L54

\bibitem[{{Bruzual} \& {Charlot}(2003)}]{bruzual03}
{Bruzual} G., {Charlot} S., 2003, \mnras, 344, 1000

\bibitem[{{Bullock} {et~al.}(2001){Bullock}, {Kolatt}, {Sigad}, {Somerville},
  {Kravtsov}, {Klypin}, {Primack}, \& {Dekel}}]{bullock01}
{Bullock} J.~S., {Kolatt} T.~S., {Sigad} Y., {Somerville} R.~S., {Kravtsov}
  A.~V., {Klypin} A.~A., {Primack} J.~R., {Dekel} A., 2001, \mnras, 321, 559

\bibitem[{{Casertano} \& {Hut}(1985)}]{casertano85}
{Casertano} S., {Hut} P., 1985, \apj, 298, 80

\bibitem[{{Chien} \& {Barnes}(2010)}]{chien10}
{Chien} L.-H., {Barnes} J.~E., 2010, \mnras, 407, 43

\bibitem[{{Chies-Santos} {et~al.}(2011){Chies-Santos}, {Larsen}, {Cantiello},
  {Strader}, {Kuntschner}, {Wehner}, \& {Brodie}}]{chies11}
{Chies-Santos} A.~L., {Larsen} S.~S., {Cantiello} M., {Strader} J.,
  {Kuntschner} H., {Wehner} E.~M., {Brodie} J.~P., 2011, \aap~submitted

\bibitem[{{Cole} {et~al.}(2000){Cole}, {Lacey}, {Baugh}, \& {Frenk}}]{cole00}
{Cole} S., {Lacey} C.~G., {Baugh} C.~M., {Frenk} C.~S., 2000, \mnras, 319, 168

\bibitem[{{de Vaucouleurs}(1948)}]{vaucouleurs48}
{de Vaucouleurs} G., 1948, Annales d'Astrophysique, 11, 247

\bibitem[{{Elmegreen}(1983)}]{elmegreen83}
{Elmegreen} B.~G., 1983, \mnras, 203, 1011

\bibitem[{{Elmegreen}(2010)}]{elmegreen10}
---, 2010, \apjl, 712, L184

\bibitem[{{Elmegreen} \& {Efremov}(1997)}]{elmegreen97}
{Elmegreen} B.~G., {Efremov} Y.~N., 1997, \apj, 480, 235

\bibitem[{{Elmegreen} \& {Hunter}(2010)}]{elmegreen10b}
{Elmegreen} B.~G., {Hunter} D.~A., 2010, \apj, 712, 604

\bibitem[{{Fall} \& {Zhang}(2001)}]{fall01}
{Fall} S.~M., {Zhang} Q., 2001, \apj, 561, 751

\bibitem[{{Forbes} {et~al.}(1997){Forbes}, {Brodie}, \& {Grillmair}}]{forbes97}
{Forbes} D.~A., {Brodie} J.~P., {Grillmair} C.~J., 1997, \aj, 113, 1652

\bibitem[{{Gerritsen} \& {Icke}(1997)}]{gerritsen97}
{Gerritsen} J.~P.~E., {Icke} V., 1997, \aap, 325, 972

\bibitem[{{Gieles}(2009)}]{gieles09}
{Gieles} M., 2009, \mnras, 394, 2113

\bibitem[{{Gieles} {et~al.}(2007){Gieles}, {Athanassoula}, \& {Portegies
  Zwart}}]{gieles07}
{Gieles} M., {Athanassoula} E., {Portegies Zwart} S.~F., 2007, \mnras, 376, 809

\bibitem[{{Gieles} \& {Baumgardt}(2008)}]{gieles08}
{Gieles} M., {Baumgardt} H., 2008, \mnras, 389, L28

\bibitem[{{Gieles} {et~al.}(2011){Gieles}, {Heggie}, \& {Zhao}}]{gieles11b}
{Gieles} M., {Heggie} D.~C., {Zhao} H., 2011, \mnras, 413, 2509

\bibitem[{{Gieles} {et~al.}(2006){Gieles}, {Portegies Zwart}, {Baumgardt},
  {Athanassoula}, {Lamers}, {Sipior}, \& {Leenaarts}}]{gieles06}
{Gieles} M., {Portegies Zwart} S.~F., {Baumgardt} H., {Athanassoula} E.,
  {Lamers} H.~J.~G.~L.~M., {Sipior} M., {Leenaarts} J., 2006, \mnras, 371, 793

\bibitem[{{Gnedin} {et~al.}(1999){Gnedin}, {Hernquist}, \&
  {Ostriker}}]{gnedin99b}
{Gnedin} O.~Y., {Hernquist} L., {Ostriker} J.~P., 1999, \apj, 514, 109

\bibitem[{{Goddard} {et~al.}(2010){Goddard}, {Bastian}, \&
  {Kennicutt}}]{goddard10}
{Goddard} Q.~E., {Bastian} N., {Kennicutt} R.~C., 2010, \mnras, 405, 857

\bibitem[{{Goodwin} \& {Bastian}(2006)}]{goodwin06}
{Goodwin} S.~P., {Bastian} N., 2006, \mnras, 373, 752

\bibitem[{{Goudfrooij} {et~al.}(2004){Goudfrooij}, {Gilmore}, {Whitmore}, \&
  {Schweizer}}]{goudfrooij04}
{Goudfrooij} P., {Gilmore} D., {Whitmore} B.~C., {Schweizer} F., 2004, \apjl,
  613, L121

\bibitem[{{Goudfrooij} {et~al.}(2007){Goudfrooij}, {Schweizer}, {Gilmore}, \&
  {Whitmore}}]{goudfrooij07}
{Goudfrooij} P., {Schweizer} F., {Gilmore} D., {Whitmore} B.~C., 2007, \aj,
  133, 2737

\bibitem[{{Harris}(1996)}]{harris96}
{Harris} W.~E., 1996, \aj, 112, 1487

\bibitem[{{Harris}(2009)}]{harris09b}
---, 2009, \apj, 703, 939

\bibitem[{{Harris} \& {Pudritz}(1994)}]{harris94}
{Harris} W.~E., {Pudritz} R.~E., 1994, \apj, 429, 177

\bibitem[{{Hernquist}(1989)}]{hernquist89}
{Hernquist} L., 1989, \nat, 340, 687

\bibitem[{{Hernquist}(1990)}]{hernquist90}
---, 1990, \apj, 356, 359

\bibitem[{{Hodge}(1961)}]{hodge61}
{Hodge} P.~W., 1961, \aj, 66, 83

\bibitem[{{Holtzman} {et~al.}(1992){Holtzman}, {Faber}, {Shaya}, {Lauer},
  {Groth}, {Hunter}, {Baum}, {Ewald}, {Hester}, {Light}, {Lynds}, {O'Neil}, \&
  {Westphal}}]{holtzman92}
{Holtzman} J.~A., {Faber} S.~M., {Shaya} E.~J., {Lauer} T.~R., {Groth} J.,
  {Hunter} D.~A., {Baum} W.~A., {Ewald} S.~P., {Hester} J.~J., {Light} R.~M.,
  {Lynds} C.~R., {O'Neil} Jr. E.~J., {Westphal} J.~A., 1992, \aj, 103, 691

\bibitem[{{Hopkins} {et~al.}(2009){Hopkins}, {Cox}, {Younger}, \&
  {Hernquist}}]{hopkins09}
{Hopkins} P.~F., {Cox} T.~J., {Younger} J.~D., {Hernquist} L., 2009, \apj, 691,
  1168

\bibitem[{{Jord{\'a}n} {et~al.}(2007){Jord{\'a}n}, {McLaughlin},
  {C{\^o}t{\'e}}, {Ferrarese}, {Peng}, {Mei}, {Villegas}, {Merritt}, {Tonry},
  \& {West}}]{jordan07}
{Jord{\'a}n} A., {McLaughlin} D.~E., {C{\^o}t{\'e}} P., {Ferrarese} L., {Peng}
  E.~W., {Mei} S., {Villegas} D., {Merritt} D., {Tonry} J.~L., {West} M.~J.,
  2007, \apjs, 171, 101

\bibitem[{{Karl} {et~al.}(2010){Karl}, {Naab}, {Johansson}, {Kotarba}, {Boily},
  {Renaud}, \& {Theis}}]{karl10}
{Karl} S.~J., {Naab} T., {Johansson} P.~H., {Kotarba} H., {Boily} C.~M.,
  {Renaud} F., {Theis} C., 2010, \apjl, 715, L88

\bibitem[{{Kennicutt}(1989)}]{kennicutt89}
{Kennicutt} Jr. R.~C., 1989, \apj, 344, 685

\bibitem[{{Kruijssen}(2009)}]{kruijssen09c}
{Kruijssen} J.~M.~D., 2009, \aap, 507, 1409

\bibitem[{{Kruijssen} \& {Cooper}(2011)}]{kruijssen11d}
{Kruijssen} J.~M.~D., {Cooper} A.~P., 2011, \mnras~in press, {{\tt
  ArXiV:1110.4106}}

\bibitem[{{Kruijssen} \& {Lamers}(2008)}]{kruijssen08}
{Kruijssen} J.~M.~D., {Lamers} H.~J.~G.~L.~M., 2008, \aap, 490, 151

\bibitem[{{Kruijssen} {et~al.}(2011{\natexlab{a}}){Kruijssen}, {Maschberger},
  {Moeckel}, {Clarke}, {Bastian}, \& {Bonnell}}]{kruijssen11b}
{Kruijssen} J.~M.~D., {Maschberger} T., {Moeckel} N., {Clarke} C.~J., {Bastian}
  N., {Bonnell} I.~A., 2011{\natexlab{a}}, \mnras~in press,~{{\tt
  ArXiV:1109.0986}}

\bibitem[{{Kruijssen} {et~al.}(2011{\natexlab{b}}){Kruijssen}, {Pelupessy},
  {Lamers}, {Portegies Zwart}, \& {Icke}}]{kruijssen11}
{Kruijssen} J.~M.~D., {Pelupessy} F.~I., {Lamers} H.~J.~G.~L.~M., {Portegies
  Zwart} S.~F., {Icke} V., 2011{\natexlab{b}}, \mnras, 414, 1339

\bibitem[{{Kruijssen} \& {Portegies Zwart}(2009)}]{kruijssen09b}
{Kruijssen} J.~M.~D., {Portegies Zwart} S.~F., 2009, \apjl, 698, L158

\bibitem[{{Kundu} \& {Whitmore}(2001)}]{kundu01}
{Kundu} A., {Whitmore} B.~C., 2001, \aj, 121, 2950

\bibitem[{{K{\"u}pper} {et~al.}(2008){K{\"u}pper}, {Kroupa}, \&
  {Baumgardt}}]{kuepper08}
{K{\"u}pper} A.~H.~W., {Kroupa} P., {Baumgardt} H., 2008, \mnras, 389, 889

\bibitem[{{Lada} \& {Lada}(2003)}]{lada03}
{Lada} C.~J., {Lada} E.~A., 2003, \araa, 41, 57

\bibitem[{{Lamers} {et~al.}(2010){Lamers}, {Baumgardt}, \& {Gieles}}]{lamers10}
{Lamers} H.~J.~G.~L.~M., {Baumgardt} H., {Gieles} M., 2010, \mnras, 409, 305

\bibitem[{{Lamers} \& {Gieles}(2006)}]{lamers06a}
{Lamers} H.~J.~G.~L.~M., {Gieles} M., 2006, \aap, 455, L17

\bibitem[{{Lamers} {et~al.}(2005{\natexlab{a}}){Lamers}, {Gieles}, {Bastian},
  {Baumgardt}, {Kharchenko}, \& {Portegies Zwart}}]{lamers05}
{Lamers} H.~J.~G.~L.~M., {Gieles} M., {Bastian} N., {Baumgardt} H.,
  {Kharchenko} N.~V., {Portegies Zwart} S., 2005{\natexlab{a}}, \aap, 441, 117

\bibitem[{{Lamers} {et~al.}(2005{\natexlab{b}}){Lamers}, {Gieles}, \&
  {Portegies Zwart}}]{lamers05a}
{Lamers} H.~J.~G.~L.~M., {Gieles} M., {Portegies Zwart} S.~F.,
  2005{\natexlab{b}}, \aap, 429, 173

\bibitem[{{Larsen}(2004)}]{larsen04b}
{Larsen} S.~S., 2004, \aap, 416, 537

\bibitem[{{Larsen}(2009)}]{larsen09}
---, 2009, \aap, 494, 539

\bibitem[{{Larsen} {et~al.}(2001){Larsen}, {Brodie}, {Huchra}, {Forbes}, \&
  {Grillmair}}]{larsen01}
{Larsen} S.~S., {Brodie} J.~P., {Huchra} J.~P., {Forbes} D.~A., {Grillmair}
  C.~J., 2001, \aj, 121, 2974

\bibitem[{{Li} {et~al.}(2004){Li}, {Mac Low}, \& {Klessen}}]{li04}
{Li} Y., {Mac Low} M., {Klessen} R.~S., 2004, \apjl, 614, L29

\bibitem[{{Marigo} {et~al.}(2008){Marigo}, {Girardi}, {Bressan}, {Groenewegen},
  {Silva}, \& {Granato}}]{marigo08}
{Marigo} P., {Girardi} L., {Bressan} A., {Groenewegen} M.~A.~T., {Silva} L.,
  {Granato} G.~L., 2008, \aap, 482, 883

\bibitem[{{Mihos} \& {Hernquist}(1996)}]{mihos96}
{Mihos} J.~C., {Hernquist} L., 1996, \apj, 464, 641

\bibitem[{{Miller} {et~al.}(1997){Miller}, {Whitmore}, {Schweizer}, \&
  {Fall}}]{miller97}
{Miller} B.~W., {Whitmore} B.~C., {Schweizer} F., {Fall} S.~M., 1997, \aj, 114,
  2381

\bibitem[{{Mo} {et~al.}(1998){Mo}, {Mao}, \& {White}}]{mo98}
{Mo} H.~J., {Mao} S., {White} S.~D.~M., 1998, \mnras, 295, 319

\bibitem[{{Monaghan}(1992)}]{monaghan92}
{Monaghan} J.~J., 1992, \araa, 30, 543

\bibitem[{{Muratov} \& {Gnedin}(2010)}]{muratov10}
{Muratov} A.~L., {Gnedin} O.~Y., 2010, \apj, 718, 1266

\bibitem[{{Pelupessy}(2005)}]{pelupessy05}
{Pelupessy} F.~I., 2005, PhD thesis, Leiden Observatory, Leiden University,
  P.O.~Box 9513, 2300 RA Leiden, The Netherlands

\bibitem[{{Pelupessy} \& {Papadopoulos}(2009)}]{pelupessy09}
{Pelupessy} F.~I., {Papadopoulos} P.~P., 2009, \apj, 707, 954

\bibitem[{{Pelupessy} {et~al.}(2006){Pelupessy}, {Papadopoulos}, \& {van der
  Werf}}]{pelupessy06}
{Pelupessy} F.~I., {Papadopoulos} P.~P., {van der Werf} P., 2006, \apj, 645,
  1024

\bibitem[{{Pelupessy} \& {Portegies Zwart}(2011)}]{pelupessy11}
{Pelupessy} F.~I., {Portegies Zwart} S.~F., 2011, \mnras~accepted, {{\tt
  ArXiv:1111.0992}}

\bibitem[{{Pelupessy} {et~al.}(2004){Pelupessy}, {van der Werf}, \&
  {Icke}}]{pelupessy04}
{Pelupessy} F.~I., {van der Werf} P.~P., {Icke} V., 2004, \aap, 422, 55

\bibitem[{{Peng} {et~al.}(2006){Peng}, {C{\^o}t{\'e}}, {Jord{\'a}n},
  {Blakeslee}, {Ferrarese}, {Mei}, {West}, {Merritt}, {Milosavljevi{\'c}}, \&
  {Tonry}}]{peng06}
{Peng} E.~W., {C{\^o}t{\'e}} P., {Jord{\'a}n} A., {Blakeslee} J.~P.,
  {Ferrarese} L., {Mei} S., {West} M.~J., {Merritt} D., {Milosavljevi{\'c}} M.,
  {Tonry} J.~L., 2006, \apj, 639, 838

\bibitem[{{Peng} {et~al.}(2008){Peng}, {Jord{\'a}n}, {C{\^o}t{\'e}},
  {Takamiya}, {West}, {Blakeslee}, {Chen}, {Ferrarese}, {Mei}, {Tonry}, \&
  {West}}]{peng08}
{Peng} E.~W., {Jord{\'a}n} A., {C{\^o}t{\'e}} P., {Takamiya} M., {West} M.~J.,
  {Blakeslee} J.~P., {Chen} C.-W., {Ferrarese} L., {Mei} S., {Tonry} J.~L.,
  {West} A.~A., 2008, \apj, 681, 197

\bibitem[{{Portegies Zwart} {et~al.}(1998){Portegies Zwart}, {Hut}, {Makino},
  \& {McMillan}}]{portegieszwart98}
{Portegies Zwart} S.~F., {Hut} P., {Makino} J., {McMillan} S.~L.~W., 1998,
  \aap, 337, 363

\bibitem[{{Portegies Zwart} {et~al.}(2010){Portegies Zwart}, {McMillan}, \&
  {Gieles}}]{portegieszwart10}
{Portegies Zwart} S.~F., {McMillan} S.~L.~W., {Gieles} M., 2010, \araa, 48, 431

\bibitem[{{Praagman} {et~al.}(2010){Praagman}, {Hurley}, \&
  {Power}}]{praagman10}
{Praagman} A., {Hurley} J., {Power} C., 2010, New Ast., 15, 46

\bibitem[{{Prieto} \& {Gnedin}(2008)}]{prieto08}
{Prieto} J.~L., {Gnedin} O.~Y., 2008, \apj, 689, 919

\bibitem[{{Qu} {et~al.}(2011){Qu}, {Di Matteo}, {Lehnert}, {van Driel}, \&
  {Jog}}]{qu11}
{Qu} Y., {Di Matteo} P., {Lehnert} M.~D., {van Driel} W., {Jog} C.~J., 2011,
  ArXiv e-prints

\bibitem[{{Renaud} {et~al.}(2011){Renaud}, {Gieles}, \& {Boily}}]{renaud11}
{Renaud} F., {Gieles} M., {Boily} C., 2011, \mnras~in press, {{\tt
  ArXiV:1107.5820}}

\bibitem[{{Schechter}(1976)}]{schechter76}
{Schechter} P., 1976, \apj, 203, 297

\bibitem[{{Schmidt}(1959)}]{schmidt59}
{Schmidt} M., 1959, \apj, 129, 243

\bibitem[{{Schweizer}(1982)}]{schweizer82}
{Schweizer} F., 1982, \apj, 252, 455

\bibitem[{{Schweizer}(1987)}]{schweizer87}
---, 1987, in Nearly Normal Galaxies. From the Planck Time to the Present, New
  York, Springer-Verlag, {S.~M.~Faber}, ed., pp. 18--25

\bibitem[{{Schweizer} {et~al.}(1996){Schweizer}, {Miller}, {Whitmore}, \&
  {Fall}}]{schweizer96}
{Schweizer} F., {Miller} B.~W., {Whitmore} B.~C., {Fall} S.~M., 1996, \aj, 112,
  1839

\bibitem[{{Schweizer} \& {Seitzer}(1998)}]{schweizer98}
{Schweizer} F., {Seitzer} P., 1998, \aj, 116, 2206

\bibitem[{{Searle} \& {Zinn}(1978)}]{searle78}
{Searle} L., {Zinn} R., 1978, \apj, 225, 357

\bibitem[{{Shapiro} {et~al.}(2010){Shapiro}, {Genzel}, \& {F{\"o}rster
  Schreiber}}]{shapiro10}
{Shapiro} K.~L., {Genzel} R., {F{\"o}rster Schreiber} N.~M., 2010, \mnras, 403,
  L36

\bibitem[{{Shapley}(1939)}]{shapley39}
{Shapley} H., 1939, Proceedings of the National Academy of Science, 25, 565

\bibitem[{{Silva-Villa} \& {Larsen}(2011)}]{silvavilla11}
{Silva-Villa} E., {Larsen} S.~S., 2011, \aap, 529, A25+

\bibitem[{{Spitzer}(1987)}]{spitzer87}
{Spitzer} L., 1987, {Dynamical evolution of globular clusters}. Princeton, NJ,
  Princeton University Press, 1987, 191 p.

\bibitem[{{Spitzer}(1958)}]{spitzer58}
{Spitzer} Jr. L., 1958, \apj, 127, 17

\bibitem[{{Springel}(2010)}]{springel10}
{Springel} V., 2010, \mnras, 401, 791

\bibitem[{{Springel} {et~al.}(2005){Springel}, {Di Matteo}, \&
  {Hernquist}}]{springel05b}
{Springel} V., {Di Matteo} T., {Hernquist} L., 2005, \mnras, 361, 776

\bibitem[{{Springel} \& {Hernquist}(2002)}]{springel02}
{Springel} V., {Hernquist} L., 2002, \mnras, 333, 649

\bibitem[{{Strader} {et~al.}(2011){Strader}, {Romanowsky}, {Brodie}, {Spitler},
  {Beasley}, {Arnold}, {Tamura}, {Sharples}, \& {Arimoto}}]{strader11}
{Strader} J., {Romanowsky} A., {Brodie} J., {Spitler} L., {Beasley} M.,
  {Arnold} J., {Tamura} N., {Sharples} R., {Arimoto} N., 2011, \apjs~in~press,
  {{\tt ArXiV:1110.2778}}

\bibitem[{{Tanikawa} \& {Fukushige}(2010)}]{tanikawa10}
{Tanikawa} A., {Fukushige} T., 2010, \pasj, 62, 1215

\bibitem[{{Vesperini}(2001)}]{vesperini01}
{Vesperini} E., 2001, \mnras, 322, 247

\bibitem[{{Vesperini} \& {Heggie}(1997)}]{vesperini97b}
{Vesperini} E., {Heggie} D.~C., 1997, \mnras, 289, 898

\bibitem[{{Vesperini} {et~al.}(2003){Vesperini}, {Zepf}, {Kundu}, \&
  {Ashman}}]{vesperini03}
{Vesperini} E., {Zepf} S.~E., {Kundu} A., {Ashman} K.~M., 2003, \apj, 593, 760

\bibitem[{{Weinberg}(1994)}]{weinberg94b}
{Weinberg} M.~D., 1994, \aj, 108, 1403

\bibitem[{{White} \& {Frenk}(1991)}]{white91}
{White} S.~D.~M., {Frenk} C.~S., 1991, \apj, 379, 52

\bibitem[{{White} \& {Rees}(1978)}]{white78}
{White} S.~D.~M., {Rees} M.~J., 1978, \mnras, 183, 341

\bibitem[{{Whitmore} {et~al.}(2007){Whitmore}, {Chandar}, \&
  {Fall}}]{whitmore07}
{Whitmore} B.~C., {Chandar} R., {Fall} S.~M., 2007, \aj, 133, 1067

\bibitem[{{Whitmore} {et~al.}(1999){Whitmore}, {Zhang}, {Leitherer}, {Fall},
  {Schweizer}, \& {Miller}}]{whitmore99}
{Whitmore} B.~C., {Zhang} Q., {Leitherer} C., {Fall} S.~M., {Schweizer} F.,
  {Miller} B.~W., 1999, \aj, 118, 1551

\bibitem[{{Yoon} {et~al.}(2011){Yoon}, {Lee}, {Blakeslee}, {Peng}, {Sohn},
  {Cho}, {Kim}, {Chung}, {Kim}, \& {Lee}}]{yoon11}
{Yoon} S.-J., {Lee} S.-Y., {Blakeslee} J.~P., {Peng} E.~W., {Sohn} S.~T., {Cho}
  J., {Kim} H.-S., {Chung} C., {Kim} S., {Lee} Y.-W., 2011, \apj~in press,
  {{\tt ArXiv:1109.5178}}

\bibitem[{{Yoon} {et~al.}(2006){Yoon}, {Yi}, \& {Lee}}]{yoon06}
{Yoon} S.-J., {Yi} S.~K., {Lee} Y.-W., 2006, Science, 311, 1129

\bibitem[{{Zepf} {et~al.}(1999){Zepf}, {Ashman}, {English}, {Freeman}, \&
  {Sharples}}]{zepf99}
{Zepf} S.~E., {Ashman} K.~M., {English} J., {Freeman} K.~C., {Sharples} R.~M.,
  1999, \aj, 118, 752

\bibitem[{{Zhang} \& {Fall}(1999)}]{zhang99}
{Zhang} Q., {Fall} S.~M., 1999, \apjl, 527, L81

\end{thebibliography}

\bsp

\label{lastpage}

\end{document}